\documentclass[a4paper,11pt]{article}
\usepackage{jheppub} 
\usepackage{lineno}
\usepackage{slashed}

\title{Majorana Neutrino Physics and Matter-Antimatter Asymmetry in Type IIA String Theory}

\author[1]{Yang Liu} 
\affiliation[1]{Department of Physics, Tsinghua University, Beijing 100084, China}
\emailAdd{liu-yang\_1990@mail.tsinghua.edu.cn}

\abstract{We study Majorana neutrino masses and the matter–antimatter asymmetry in Type IIA string theory on a $T^6/(\mathbb{Z}_2 \times \mathbb{Z}_2)$ orientifold with intersecting D6-branes. Euclidean D2-instantons with $\text{Sp}(2)$ Chan-Paton symmetry generate both the seesaw mechanism and the Weinberg operator. The flavour vectors are not computed from a compactification but fixed by matching, so our results are an existence proof. Matching the measured mass-squared differences selects the seesaw-dominated case with normal ordering. Swampland criteria constrain the instanton action and the reheating temperature. A benchmark  example reproduces the neutrino parameters, gives $\eta_B \approx 8.6 \times 10^{-11}$, within an order of magnitude of observation, and predicts $\langle m_{\beta \beta} \rangle =2.594 \text{meV}$. A parameter scan shows the mass hierarchy is accommodated rather than predicted.}

\begin{document}
\maketitle
\flushbottom

\section{Introduction}
\label{sec:intro}

The Standard Model (SM) of particle physics has achieved remarkable success in describing electromagnetic, weak, and strong interactions up to the TeV scale. However, it leaves several fundamental problems unresolved, which point to the necessity of physics beyond the Standard Model (BSM). Among these, two key issues are particularly important:\\
\begin{itemize}
 \item Neutrino Physics: The observation of neutrino oscillations provides the first definitive evidence of physics beyond the Standard Model, requiring nonzero neutrino masses and lepton flavor violation. While the Standard Model predicts strictly massless neutrinos, the tiny yet nonzero masses point toward new mechanisms, such as the seesaw mechanism or higher-dimensional operators that often involve heavy right-handed neutrinos and new symmetry scales. Understanding the origin of these masses remains a central open question in particle physics and cosmology \cite{fukuda1998evidence,ahmad2002direct}.
 \item Baryon Asymmetry: The observed matter-antimatter asymmetry, quantified by the baryon-to-photon ratio $\eta_B \sim 6.1 \times 10^{-10}$, cannot be explained within the Standard Model, which lacks sufficient CP violation and fails to provide a strong first-order electroweak phase transition. Satisfying Sakharov's three conditions - baryon number violation, C and CP violation, and departure from thermal equilibrium - requires new physics beyond the Standard Model, often connected to leptogenesis, electroweak baryogenesis, or Affleck-Dine mechanisms. Resolving this asymmetry is a fundamental challenge linking high-energy physics with early-universe cosmology \cite{covi1996cp,giudice2004towards}.
\end{itemize}

String theory is a leading candidate for a unified framework, therefore, it should address these problems simultaneously.
In this work, we focus on Type IIA string theory compactified on a $T^6/(\mathbb{Z}_2 \times \mathbb{Z}_2)$ orientifold with intersecting D6-branes. The intersecting D6-brane configuration generates chiral matter and gauge interactions of the SM, while the orientifold projection guarantees the consistency of the construction by canceling anomalies. This makes Type IIA orientifold model a self-consistent framework for addressing SM shortcomings. In \cite{Liu:2025zjs}, we studied a concrete globally consistent Type IIA intersecting D6-brane model, which claimed that CP violation, matter-antimatter asymmetry, hierarchy problem, and the existence of dS vacua could be addressed in this unified model simultaneously. The goal of \cite{Liu:2025zjs} is very ambitious. But this work is more of a discussion regarding conceptual possibility; further physical details are not discussed in \cite{Liu:2025zjs}. In the present paper, we will follow the spirit of framework in \cite{Liu:2025zjs} and explore more details of Majorana neutrino physics and matter-antimatter asymmetry in Type IIA intersecting D6-brane model.

A mechanism for generating Majorana masses for neutrinos within the framework of string theory was proposed in \cite{ibanez2007neutrino,blumenhagen2007spacetime,cvetivc2007computation,ibanez2007instanton}. In this setup, certain string instanton effects can induce right-handed neutrino masses via operators of the form
\begin{equation} \label{rhnm}  
e^{-U_M} \nu_R \nu_R M_s,
\end{equation}
where $M_s$ denotes the string scale and $U_M$ is a complex scalar modulus field. Its axion-like imaginary part $\text{Im} U_M$ transforms under a gauged $U(1)_{B-L}$ symmetry, ensuring that the operator \eqref{rhnm} remains gauge invariant under $U(1)_{B-L}$. The resulting mass scale is of order $e^{-\text{Re} U_M} M_s$. Unlike conventional instanton effects -- such as those in the electroweak sector, which are suppressed as $e^{-1/\alpha_2}$, these stringy instantons need not be heavily suppressed: $\text{Re} U_M$ is not inversely related to any Standard Model gauge coupling and can be relatively small. Consequently, right-handed neutrino masses can be large, as required by phenomenological considerations. Furthermore it was observed in \cite{ibanez2007instanton} that analogous instantons can also generate a dimension 5 Weinberg operator of the form
\begin{equation} \label{d5Wo}  
e^{-U_M} \frac{1}{M_s} \bar{H} L \bar{H} L.
\end{equation}
Once the Higgs scalar acquires a vacuum expectation value, this term directly produces masses for left-handed neutrinos. Both types of instanton effects occur only in a restricted class of string compactifications, where the Standard Model gauge group is extended by a $U(1)_{B-L}$ gauge boson that acquires a mass via a Stückelberg mechanism. More discussions of string compactifications realizing this instanton mechanism can be found in \cite{ibanez2007neutrino,blumenhagen2007spacetime,cvetivc2007computation,ibanez2007instanton}.

In this paper, we carry out a phenomenological study of the structure of neutrino masses and mixing angles arising from the string instanton mechanism described above, using recent experimental data as a guide. Our analysis focuses on a specific class of instantons -- those with an internal $\text{Sp}(2)$ Chan-Paton (CP) symmetry -- since they exhibit the simplest structure and appear most frequently in existing instanton searches \cite{ibanez2007instanton}. For such instantons, the flavor dependence of both the right-handed neutrino masses and the Weinberg operator factorizes into products of flavor vectors (denoted as $d_a$ and $c_a$ for $a=1,2,3$ in the main text for the $\nu_R$-mass term and the Weinberg operator, respectively). In principle, these flavor vectors $d_a$ and $c_a$ can be computed from the details of the underlying string compactification. This simple flavor structure, together with the expectation that multiple distinct instantons contribute to the amplitudes, makes it natural to obtain a hierarchical pattern of neutrino masses \cite{ibanez2007instanton}. In the present work we do not carry out this computation. We instead treat the light neutrino masses, mixing angles and right-handed neutrino masses as inputs, and use the instanton relations to determine what the flavour vectors and instanton actions would have to be. Our results should therefore be read as an existence proof -- a demonstration that the Type IIA instanton mechanism can be made quantitatively consistent with both neutrino data and the baryon asymmetry -- and not as a first-principles derivation of either. We indicate at each stage which quantities are inputs and which are determined.

The discovery of neutrino oscillations proves that neutrinos have small but nonzero masses, providing the first clear evidence of physics beyond the Standard Model. The most natural explanation for their smallness is the seesaw mechanism, which introduces heavy right-handed neutrinos with Majorana masses at some high scale. Crucially, the decays of these heavy neutrinos in the early universe can violate lepton number and, through CP-violating phases in the neutrino Yukawa sector, generate a net lepton asymmetry. This CP asymmetry $\epsilon_1$ in the decays of the lightest right-handed neutrino is directly related to the low-energy neutrino parameters, such as the Dirac and Majorana CP phases, linking the high-energy dynamics of leptogenesis to the observable CP violation in neutrino oscillations. The resulting lepton asymmetry is then partially converted into a baryon asymmetry by $(B+L)$-violating sphaleron processes, which are active at temperatures above the electroweak scale. The final baryon asymmetry is given by
\begin{equation} \label{fba1}  
\eta_B  \approx \frac{28}{79} \frac{n^{eq}_N}{s} \epsilon_1 \kappa_f \approx 0.01 \epsilon_1 \kappa_f \sim 6.1 \times 10^{-10},
\end{equation}
where $\kappa_f$ encodes washout effects. Thus, the observed matter-antimatter asymmetry can be traced back to the same neutrino sector that explains the tiny neutrino masses, providing a deep and elegant connection between neutrino physics, CP violation, and the origin of the baryon asymmetry of the universe \cite{covi1996cp,giudice2004towards}.

Dirac neutrino physics provides an instructive contrast to the Majorana scenario \cite{king2024toward}. The minimal Dirac extension requires Yukawa couplings $y_{\nu} \sim 10^{-12}$ to explain sub-eV masses -- unnatural despite technical naturalness. More elaborate Dirac seesaws impose a global $U(1)_L$ (or a discrete $\mathbb{Z}_2$) to forbid Majorana masses, and the light neutrino mass arises from a suppressed Dirac structure. In our Type IIA framework, by contrast, the D2-instanton carries $B-L$ charge $+2$ and generates the $\nu_R \nu_R$ operator directly, making neutrinos inherently Majorana. This is a sharp distinction: our mechanism predicts lepton‑number violation, while Dirac models must forbid it. Moreover, heterotic string constructions often obstruct the minimal seesaw, whereas Type IIA intersecting D6‑branes explicitly realize the instanton-induced Majorana mass. A phenomenological comparison with Dirac, radiative, Type-II seesaw and inverse‑seesaw mechanisms is given in Section 4.

While the D2-instanton generation of Majorana neutrino masses was established in the pioneering works \cite{ibanez2007neutrino,blumenhagen2007spacetime,cvetivc2007computation,ibanez2007instanton}, the present paper provides the first realization of this mechanism within a Type IIA intersecting D6-brane vacuum that simultaneously addresses neutrino masses, the baryon asymmetry, and swampland consistency. This unified framework is precisely the kind of “geometry that fits not only Einstein's equations but also the particle spectrum we want” that Vafa has emphasized as the central goal of string phenomenology \cite{Found:A}.

This paper is organized as follows: in Section 2, we review the related background of neutrino physics and matter-antimatter asymmetry. In Section 3, we discuss the theoretical mechanism of left-handed neutrino masses and mixing angles. Based on the observational data, we can determine that the mass ordering of left-handed neutrinos should be normal ordering. In Section 4, we compare the phenomenology we study in this paper with other neutrino mechanisms. In Section 5, we explore the constraints on neutrino physics from swampland conjecture. In Section 6, we show a numerical example that can be simultaneously consistent with the observational and experimental data of neutrino physics and baryon asymmetry. In Section 7, we scan the parameter space. We summarize the results that we obtain in this paper in Section 8. The appendix shows the details of calculations of our numerical example listed in Section 6.

\section{Background}

\subsection{Majorana neutrino physics}
The generation of small, non-zero neutrino masses presents a key challenge beyond the Standard Model (SM). Within the framework of string theory, particularly in certain classes of semi-realistic compactifications, a novel mechanism for generating Majorana neutrino masses has been identified. This mechanism relies on non-perturbative effects from Euclidean D-branes, or “string instantons”, which can violate global symmetries (like lepton number) that are intact at the perturbative level. In this section we provide a detailed exposition of how such instantons operate within specific string constructions to yield the primary ingredients for neutrino mass generation: either direct dimension-5 Weinberg operators or large Majorana masses for right-handed neutrinos, thereby enabling the seesaw mechanism \cite{antusch2007neutrino}. 

A crucial prerequisite for this mechanism is the presence of a $U(1)_{B-L}$ gauge symmetry in the low-energy effective theory derived from the string compactification. This symmetry is anomaly-free in the presence of three right-handed neutrinos $(\nu^a_R)$ and appears naturally in many MSSM-like string models. However, for Majorana masses (which carry $B-L$ charge $+2$) to be allowed, this $U(1)_{B-L}$ must be broken. In the considered constructions, this occurs not via a Higgs mechanism but through a Stückelberg mechanism. A crucial coupling of the form $B \wedge F_{B-L}$ (where $B$ is a 2-form gauge field) gives a string-scale mass to the  $U(1)_{B-L}$ gauge boson. The pseudoscalar field $\eta$ dual to $B$ transforms under a $U(1)_{B-L}$ gauge transformation as $\eta(x) \rightarrow \eta(x) + q \Lambda(x)$. Consequently, from a low-energy perspective below the string scale, the gauge group is effectively just that of the SM (or a GUT extension), but the global remnant of $U(1)_{B-L}$ remains pivotal for understanding allowed operators.

The central actors are Euclidean D2-branes (D2-instantons) wrapping 3-cycles ($\Pi_M$) in the internal compact dimensions. Their classical action is given by the volume of the wrapped cycle and an axionic contribution:
\begin{equation} \label{SD2}  
S_{D2}= \frac{V_{\Pi}}{\lambda} + i \sum_r q_{M,r} \eta_r \equiv U_{\text{inst}},
\end{equation}
where $U_{\text{inst}}$ is a complex modulus field in Type IIA string thoery with intersecting D6-branes, $V_{\Pi}$ is the 3-volume wrapped by the D2-brane and $\lambda$ is the string dilaton. The imaginary part, $\text{Im} U_{\text{inst}}$, transforms under $U(1)_{B-L}$ as $\text{Im} U_{\text{inst}} \rightarrow \text{Im} U_{\text{inst}} - 2\Lambda(x)$. This transformation property is key: the exponential factor $e^{- U_{\text{inst}}}$ carries an effective $B-L$ charge of $+2$, making the operator $e^{- U_{\text{inst}}} \nu_R \nu_R$ gauge invariant \cite{antusch2007neutrino}.

The instanton can intersect the D6-branes on which the SM fields live. At the intersections between the D2-instanton and the D6-branes hosting the right-handed neutrinos, fermionic zero modes $\alpha_i$ and $\gamma_j$ ($i,j=1,2$) arise. These are Grassmann variables integrated over in the instanton path integral. They couple to the right-handed neutrinos via a superpotential interaction:
\begin{equation} \label{Sinst}  
S_{\text{inst}}(\alpha, \gamma)= d^{ij}_a (\alpha_i \nu^a \gamma_j), \quad a=1,2,3,
\end{equation}
where $d^{ij}_a$ are flavor-dependent coupling coefficients calculable from worldsheet correlators and dependent on Kähler moduli.

To generate a bilinear superpotential term for $\nu_R$, one must integrate over these fermionic zero modes:
\begin{equation} \label{ifzm}  
\int d^2 \alpha d^2 \gamma e^{-d^{ij}_a (\alpha_i \nu^a \gamma_j)} \propto \nu_a \nu_b (\epsilon_{ij} \epsilon_{kl} d^{ik}_a d^{jl}_b).
\end{equation}
Combining this with the classical instanton suppression factor yields the right-handed neutrino Majorana mass matrix:
\begin{equation} \label{rhMmm1}  
M^R_{ab} = M_s (\epsilon_{ij} \epsilon_{kl} d^{ik}_a d^{jl}_b) e^{-U_{\text{inst}}}.
\end{equation}

A significant simplification occurs for the most common class of instantons, those with an $\text{Sp}(2) = \text{SU}(2)$ Chan-Paton symmetry acting on the zero modes. In this case, the couplings factorize: $d^{ij}_a = d_a \epsilon^{ij}$, where $d_a$ is a flavor vector. The mass matrix then takes a remarkably simple, factorized form:
\begin{equation} \label{rhMmm2}  
M^R_{ab} = 2 M_s \sum_r d^{(r)}_a d^{(r)}_b  e^{-U_r}.
\end{equation}
Here, $a,b=1,2,3$. The sum runs over different instantons ($r$) that may contribute, each with its own action $U_r$ and flavor vector $d^{(r)}_a$. This structure can be written as:
\begin{equation} \label{rhMmm3}
M^R \sim  \sum_r e^{-U_r} \text{diag} (d^{(r)}_1, d^{(r)}_2, d^{(r)}_3) \cdot
\left[ \begin{array}{ccc}		
1 & 1 & 1  \\		
1 & 1 & 1  \\		
1 & 1 & 1 \\		
\end{array}	\right]
\cdot \text{diag} (d^{(r)}_1, d^{(r)}_2, d^{(r)}_3).
\end{equation}
Each instanton gives a mass to one specific linear combination of $\nu_R$'s. A hierarchy among the three right-handed neutrino masses emerges naturally if the different instantons have hierarchically different actions ($\text{Re} U_1 \gg \text{Re} U_2 \gg \text{Re} U_3$), as is generically expected since $\text{Re} U_r$ is proportional to the wrapped volume and not tied to any small SM coupling constant \cite{antusch2007neutrino}.

The seesaw mechanism can generate neutrino masses.  Once (large) right-handed neutrino masses are generated the standard see-saw mechanism is expected to induce Majorana masses for the lightest eigenvalues in the usual way, i.e. neutrino masses of the form:
\begin{equation} \label{MLss}  
M^L_{ab}({\text{see-saw}})= \frac{\langle \bar{H} \rangle^2}{2 M_s} h^T_D \left(\sum_r d^{(r)}_a d^{(r)}_b e^{-S_r} \right)^{-1} h_D,
\end{equation}
where $h_D$ is the ordinary Yukawa coupling constant in $h^{ab}_D (\nu_R \bar{H} L^b)$. The eigenvalues of these matrices are the ones which should be compared with experiment \cite{antusch2007neutrino}.

An analogous instanton mechanism can generate the dimension-5 Weinberg operator $(L^a \bar{H}) (L^b \bar{H})/M_s$ directly, without invoking the seesaw mechanism. In this case, the relevant D2-instantons have different fermionic zero modes $\beta_i$, $\delta_i$ that couple to the lepton-Higgs combination:
\begin{equation} \label{Sinstbd}  
S_{\text{inst}}(\beta, \alpha)= c^{ij}_a (\beta_i (L^a \bar{H} \delta_j)).
\end{equation}
Again, assuming an $\text{SU}(2)$ symmetry ($c^{ij}_a=c_a \epsilon^{ij}$) and integrating over the zero modes, one obtains the left-handed neutrino mass matrix after electroweak symmetry breaking:
\begin{equation} \label{MLdim5}  
M^L_{ab}= \frac{2\langle \bar{H} \rangle^2}{M_s} \sum_r c^{(r)}_a c^{(r)}_b e^{-U_r}.
\end{equation}
This matrix has an identical factorized flavor structure as $M_R$, governed by a different set of flavor vectors $c^{(r)}_a$ and instanton actions $U_r$. It provides a direct origin for small neutrino masses, competing with or supplementing the seesaw contribution \cite{antusch2007neutrino}.

The predictions of the model depend on:
\begin{itemize}
\item The string scale ($M_s$): Could range from TeV to the GUT scale. 
\item Instanton actions ($\text{Re} U_r$): These are not exponentially suppressed like QCD instantons, making the effects potentially significant. Hierarchical $\text{Re} U_r$ naturally leads to hierarchical masses.
\item Flavor vectors ($c^{(r)}_a$, $d^{(r)}_a$): These are in principle computable amplitudes from string worldsheet correlators, analogous to Yukawa couplings. They depend on Kähler moduli and brane geometry. In simple toroidal models, they relate to Jacobi theta functions, which can vanish at symmetric points, leading to texture zeros. We do not evaluate them for a specific compactification in this work; their required magnitudes are determined phenomenologically in Section 6.
\end{itemize}
It is important to note that in this stringy context, the Weinberg operator generated directly by instantons and the one obtained by integrating out the $\nu_R$'s via the seesaw mechanism are distinct contributions originating from different instantons \cite{antusch2007neutrino}. 

Therefore, in this section we have reviewed a concrete and compelling string-theoretic framework for neutrino mass generation. The mechanism is robust in a class of models with a massive $U(1)_{B-L}$. The resulting mass matrices exhibit a factorized flavor structure dictated by instanton calculus, naturally accommodate mass hierarchies, and can originate from either a direct Weinberg operator or a seesaw mechanism with heavy $\nu_R$'s, or a combination of both. The flavor structure is encoded in potentially calculable string amplitudes, setting the stage for the phenomenological analysis of mixing patterns carried out in the subsequent Section 2.2 of the paper.

\subsection{String‑theoretic origin of Majorana neutrino masses}
While deriving neutrino matrices comparable with experimental data from first principles is challenging, it is nevertheless possible in theory. In this subsection, we give a brief account of how the right-handed neutrino Majorana mass matrix emerges from Euclidean D2-brane instantons. For more comprehensive discussions, see references such as \cite{hamada2014flavor, ibanez2007neutrino, blumenhagen2007spacetime}.

In Type IIA on $T^6= T^2_1 \times T^2_2 \times T^2_3$ with intersecting D6‑branes, right‑handed neutrino fields $\nu^a_R$ arise as bifundamental chiral multiplets at the intersection of two D6-stacks, say $D6_c$ and $D6_d$.  The number of generations is the intersection number $I_{cd}= \prod^3_{i=1} I^{(i)}_{cd}$. A Majorana mass term for $\nu_R$ is forbidden at the perturbative level because it would violate the massive $U(1)_{B-L}$ gauge symmetry.  The term is instead induced non‑perturbatively by a Euclidean D2-brane (E2) instanton, denoted $D2_M$, that wraps a rigid supersymmetric 3‑cycle (one 1‑cycle on each torus).  The instanton intersects the $D6_c$ and $D6_d$ stacks, and at each intersection point there appear fermionic zero modes: a doublet $\alpha_i$ ($i=1,2$) from the $D2_M-D6_c$ intersection and a doublet $\gamma_j$ ($j=1,2$) from the $D6_d-D2_M$ intersection. The presence of exactly two $\alpha$ zero modes and two $\gamma$ zero modes is the necessary condition for the instanton to generate a $\nu_R \nu_R$ operator; if the zero-mode count were different the operator would vanish.

Integrating over the Grassmann collective coordinates $\alpha_i$ and $\gamma_j$ of the instanton yields the effective interaction
\begin{equation} \label{aginseffint}
M \int d^2 \alpha d^2 \gamma \exp[-d^{ij}_a \alpha_i \nu^a_R \gamma_j] = M \nu^a_R \nu^b_R (\epsilon_{ij} \epsilon_{kl} d^{ik}_a d^{jl}_b),
\end{equation}
where $M = M_s e^{-U_{\text{inst}}}$ and $M_s$ is the string scale.  The coefficient $d^{ij}_a$ is a three‑point coupling among the two instanton zero modes $\alpha_i$, $\gamma_j$ and the right-handed neutrino $\nu^a_R$; it is the object that encodes the flavour structure.
Because the $D2_M$ instanton wraps a product of three tori, the coupling factorizes:
\begin{equation} \label{dija123}
d^{ij}_a = d^{ij}_{a1} d^{ij}_{a2} d^{ij}_{a3}, \quad d^{ij}_{ak} (k=1,2,3)
\end{equation}
being the contribution from the $k$-th torus. When a field is localized at a single intersection point on a given torus, the corresponding flavour index is omitted. On each torus the three‑point function is computed from the worldsheet CFT as an open-string disc amplitude with boundary‑changing operators, multiplied by a one‑loop vacuum diagram; the result can be expressed in terms of Jacobi theta functions. We adopt the explicit configuration from \cite{hamada2014flavor}:
\begin{equation} \label{IcdIMcIdM}
I^{(1)}_{cd}=3, \quad I^{(2)}_{cd}=I^{(3)}_{cd}=1, \quad I_{Mc}=I_{dM}=1,
\end{equation}
so that the flavor index $a=0,1,2$ resides on the first torus. Under this choice, the three-point couplings $d^{i'j'}_a$ among the fields are given by \cite{hamada2014flavor,ibanez2007neutrino,blumenhagen2007spacetime}
\begin{equation} \label{di'j'a1}
d^{i'j'}_a = C \sum_{l \in \mathbf{Z}} \exp{\left(\frac{-A_{a i' j'}(l)}{2 \pi \alpha'} \right)},
\end{equation}
where $C$ is a flavor-independent constant arising from quantum effects, and
\begin{equation} \label{Ai'j'a}
A_{a i'j'} = \frac{1}{2} A |I_{cd} I_{dM} I_{Mc}| \left( \frac{a}{I_{cd}} + \frac{i'}{I_{dM}} + \frac{j'}{I_{Mc}} + \frac{\epsilon}{I_{dM} I_{Mc}} + l \right)^2
\end{equation}
with $A$ denoting the area of the first torus. Here, $\epsilon$ denotes the position of $D2_M$-brane on the first torus, normalized such that $\epsilon$ takes values in $[0, 1]$. Note that this coupling describes the contribution from the first torus, which determines the flavor structure; for simplicity, the index labeling that torus has been suppressed \cite{hamada2014flavor}.

Using the $\mathcal{\vartheta}$-function \cite{hamada2014flavor},
\begin{equation} \label{vartheta}
\vartheta\!\left[ \begin{matrix} a \\ b \end{matrix} \right] (\nu,\tau) = \sum_{l \in \mathbf{Z}} \exp{[\pi i (a+l)^2 \tau + 2\pi i (a+l)(\nu+b)]},
\end{equation}
the coupling can be expressed as
\begin{equation} \label{di'j'a2}
d^{i'j'}_a = C \vartheta\!\left[ \begin{matrix} \frac{a}{I_{cd}} + \frac{i'}{I_{dM}} + \frac{j'}{I_{Mc}} + \frac{\epsilon}{I_{dM} I_{Mc}} + l \\ 0 \end{matrix} \right] \left(0,\frac{iA|I_{cd} I_{dM} I_{Mc}|}{4 \pi^2 \alpha'}\right).
\end{equation}
In our specific model, we take $a=1,2,3$, $I_{cd}=3$, $i'=j'=0$, $I_{dM}=I_{Mc}=1$. The corresponding 3-point couplings then become
\begin{equation} \label{dija}
d^{ij}_a = \delta_{ij} \vartheta\!\left[ \begin{matrix} -\frac{a}{3}+\epsilon \\ 0 \end{matrix} \right] \left(0,\frac{3iA}{4 \pi^2 \alpha'} \right).
\end{equation}
Here, the indices $i$ and $j$ appearing in $\alpha_i$ and $\gamma_j$ are doublet indices under $\text{Sp}(2)$ or $\text{U}(2)$. Furthermore, using \eqref{rhMmm1}, one obtains the right-handed neutrino mass matrix.

Equation \eqref{dija} is the concrete link between the flavour vectors and the compactification data — the instanton position  and the torus area $A$. Evaluating it for a specific choice of wrapping numbers and moduli would fix $d^{(r)}_a$ from the geometry. We display it here to make the logical status of the flavour vectors explicit, but we do not evaluate it in what follows; Section 6 determines the $d^{(r)}_a$ by matching to data instead.

\subsection{Matter-antimatter asymmetry}
One of the deepest puzzles in modern cosmology and particle physics is the observed dominance of matter over antimatter in the observable universe. If matter and antimatter were exactly symmetric at early times, they would have annihilated almost completely, leaving behind only a tiny residual radiation. Instead, we observe a universe filled with baryonic matter, such as stars, galaxies, and gas, with no significant populations of antimatter domains. This net excess is quantified by the baryon asymmetry parameter:
\begin{equation} \label{bap}  
\eta_B \equiv \frac{n_B - n_{\bar{B}}}{n_{\gamma}},
\end{equation}
where $n_B$ and $n_{\bar{B}}$ are the number densities of baryons and antibaryons, respectively, and $n_{\gamma}$ is the number density of cosmic microwave background photons. This value is extracted from Big Bang Nucleosynthesis (BBN) and, with even greater precision, from Planck measurements of the cosmic microwave background anisotropies \cite{covi1996cp,giudice2004towards}.

In 1967, Andrei Sakharov formulated three necessary conditions for dynamically generating a baryon asymmetry from an initially symmetric state:\\
\begin{itemize}
    \item 1. Baryon number violation: Processes must exist that change $B \equiv n_B - n_{\bar{B}}$.
    \item 2. C and CP violation: Without charge conjugation (C) and combined charge-parity (CP) violation, equal rates for baryon-producing and antibaryon-producing reactions would cancel any net asymmetry.
    \item 3. Departure from thermal equilibrium: If all interactions are in equilibrium, CPT invariance forces the net baryon number to vanish (since the equilibrium state has zero chemical potential for conserved charges).
\end{itemize}
The Standard Model satisfies condition 1 via sphaleron processes that violate $B+L$ while conserving $B-L$, and condition 3 during the electroweak phase transition (if it is strongly first-order). However, the CP violation in the Standard Model, which arises solely from the CKM matrix, is far too small to account for the observed $\eta_B$. New physics beyond the Standard Model is therefore required \cite{covi1996cp,giudice2004towards}. 

The discovery of neutrino oscillations proves that neutrinos have small but nonzero masses, providing the first clear evidence of physics beyond the Standard Model. The most natural explanation for their smallness is the seesaw mechanism, which introduces heavy right-handed neutrinos with Majorana masses at some high scale. Crucially, the decays of these heavy neutrinos in the early universe can violate lepton number and, through CP-violating phases in the neutrino Yukawa sector, generate a net lepton asymmetry. This CP asymmetry $\epsilon_1$ in the decays of the lightest right-handed neutrino is directly related to the low-energy neutrino parameters -- including the Dirac and Majorana CP phases -- via the relation
\begin{equation} \label{e1Ynu2}  
\epsilon_1 = \frac{1}{8\pi} \frac{1}{(Y^{\dagger}_{\nu} Y_{\nu})_{11}} \sum_{j=2,3} \text{Im} [(Y^{\dagger}_{\nu} Y_{\nu})^2_{1j}] f\left(\frac{M^2_j}{M^2_1}\right)
\end{equation}
linking the high-energy dynamics of leptogenesis to the observable CP violation in neutrino oscillations. Here,
\begin{equation} \label{fMjM1}  
f(x)= \sqrt{x} [\frac{2}{1-x} + (1+x) \ln{(1 + \frac{1}{x})}-1].
\end{equation}

Among the most attractive extensions is leptogenesis, which links the baryon asymmetry to the same seesaw mechanism that explains neutrino masses. In its simplest thermal version, heavy right-handed neutrinos $N_i$ decay out of equilibrium, producing a lepton asymmetry that is partially converted into a baryon asymmetry by sphalerons. The final baryon asymmetry is given by
\begin{equation} \label{fba2}  
\eta_B \approx \frac{28}{79} \frac{n^{eq}_N}{s} \epsilon_1 \kappa_f \approx 0.01 \epsilon_1 \kappa_f,
\end{equation}
where $n^{eq}_N/s$ is the equilibrium abundance of the decaying heavy neutrino relative to entropy density, $\epsilon_1$ is the CP asymmetry in its decays (typically $\sim 10^{-6}-10^{-5}$ for viable scenarios), and $\kappa_f<1$ is an efficiency factor accounting for washout processes. The overall factor $28/79$ arises from the sphaleron conversion of lepton number to baryon number in the Standard Model \cite{covi1996cp,giudice2004towards}.

Alternatively, electroweak baryogenesis utilizes a strongly first-order electroweak phase transition together with new sources of CP violation from extended Higgs sectors (e.g., two-Higgs-doublet models or supersymmetry). The resulting baryon asymmetry can be computed from transport equations and typically involves bubble wall profiles and CP-violating source terms.

Regardless of the specific mechanism, any successful theory of baryogenesis must reproduce the observed value $\eta_B \sim 6.1 \times 10^{-10}$. This stringent requirement makes the baryon asymmetry a powerful probe of high-energy physics, intimately connecting the origin of matter with neutrino masses, CP violation, and the thermal history of the early universe \cite{covi1996cp,giudice2004towards}.

\section{Majorana neutrino masses and mixing angles}
In this section, we discuss the theoretical mechanism of left-handed neutrino masses and mixing angles. The leptonic mixing matrix, commonly referred to as the PMNS matrix, is conventionally expressed in the standard PDG parametrization as
\begin{equation} \label{UPMNS1}
U_{PMNS} = 
\left[ \begin{array}{ccc}		
c_{12} c_{13} & s_{12}c_{13} & s_{13}e^{-i \delta} \\		
-c_{23} s_{12} - s_{13} s_{23} c_{12} e^{i \delta} & c_{23} c_{12} - s_{13} s_{23} s_{12} e^{i \delta} &  s_{23} c_{13} \\		
s_{23} s_{12} - s_{13} c_{23} c_{12} e^{i \delta} & -s_{23} c_{12} - s_{13} c_{23} s_{12} e^{i \delta} & c_{23} c_{13} \\		
\end{array}	\right]
\cdot P_{\text{Maj}},
\end{equation}
where the shorthand notations $s_{ij}= \sin (\theta_{ij})$ and $c_{ij} =\cos (\theta_{ij})$ are employed. The parameter $\delta$ denotes the Dirac CP-violating phase, which can in principle be measured through neutrino oscillation experiments. Furthermore,
\begin{equation} \label{PMaj}
P_{\text{Maj}}=\text{diag} (e^{i\alpha_1/2},e^{i\alpha_2/2},1)
\end{equation}
encodes the Majorana phases $\alpha_1$ and $\alpha_2$. The overall PMNS mixing arises from two sources: the matrix $V_{eL}$ that diagonalizes the charged-lepton mass matrix, and the matrix $V_{\nu_L}$ that diagonalizes the neutrino mass matrix:
\begin{equation} \label{UPMNS2}
U_{PMNS} = V_{eL} V^{\dagger}_{\nu_L}.
\end{equation}
An equivalent representation of the PMNS matrix is
\begin{equation} \label{UPMNS3}
U_{PMNS} = 
\left[ \begin{array}{ccc}		
U_{0,e1} & U_{0,e2} & U_{0,e3} \\		
U_{0,\mu1} & U_{0,\mu2} & U_{0,\mu3} \\		
U_{0,\tau1} & U_{0,\tau2} & U_{0,\tau3}\\		
\end{array}	\right]
\cdot P_{\text{Maj}}
=\left[ \begin{array}{ccc}		
U_{e1} & U_{e2} & U_{e3} \\		
U_{\mu1} & U_{\mu2} & U_{\mu3} \\		
U_{\tau1} & U_{\tau2} & U_{\tau3}\\		
\end{array}	\right]
\end{equation}
In this work, we adopt the hypothesis that the substantial mixing observed in the lepton sector originates primarily from the neutrino mass matrix. Such large mixings are naturally anticipated within our framework, given that neutrino masses arise from string instanton effects -- a mechanism fundamentally distinct from the origin of masses for charged leptons and quarks. Under this assumption, the relatively small mixings associated with the charged-lepton mass matrix can be treated perturbatively \cite{antusch2007neutrino}.

We follow the standard PDG convention for labelling the three light Majorana mass eigenstates. The states $\nu_1$ and $\nu_2$ are the pair separated by the solar mass gap, ordered such that $|U_{e1}|^2>|U_{e2}|^2$, which given the measured value $\theta_{12}=33.5^\circ<45^\circ$ means $\nu_1$ is the more electron-rich of the two; the state $\nu_3$ is the one separated from this pair by the atmospheric gap and is characterized by $|U_{e3}|^2=\sin^2\theta_{13}\simeq0.022$. With this labelling, $\Delta m^2_{21}\equiv m_2^2-m_1^2>0$ is not a convention but an experimental result. The open question of the mass ordering is then the sign of $\Delta m^2_{31}$: whether the state with the smallest electron content, $\nu_3$, is heavier than the solar pair (normal ordering, NO) or lighter than it (inverse ordering, IO).

\subsection{Masses and mixings from the Weinberg operator}
We begin by considering the scenario in which the neutrino mass matrix is dominated by the contribution from the dimension-5 Weinberg operator. As previously mentioned, in this case the left-handed neutrino mass matrix is directly linked to the instanton contribution discussed in the preceding section.

\subsubsection{The general normal hierarchy case}
When the Weinberg operator provides the dominant contribution, the instanton-induced Majorana neutrino mass matrix can be expressed as
\begin{equation} \label{MLij}
M^L_{ij} = \sum_r I_r c^{(r)}_i c^{(r)}_j,
\end{equation}
where we define
\begin{equation} \label{Ir}
I_r = 2 \frac{\langle \bar{H} \rangle^2}{M_s} e^{-S_r}.
\end{equation}
We now focus on the scenario involving three distinct instanton contributions to the neutrino mass matrix, labeled by $r=1,2,3$. A certain hierarchy, possibly mild, is assumed among them, namely
\begin{equation} \label{hierarchy}
|I_3 c^{(3)}_i c^{(3)}_j| \gg |I_2 c^{(2)}_i c^{(2)}_j| \gg |I_1 c^{(1)}_i c^{(1)}_j|.
\end{equation}
This hierarchy can be motivated by a corresponding ordering of the instanton factors,
\begin{equation} \label{hierarchyinstanton}
e^{-\text{Re}(S_3)} \gg e^{-\text{Re}(S_2)} \gg e^{-\text{Re}(S_1)}.
\end{equation}
As discussed in the previous section, such moderate hierarchies are plausible in orientifold compactifications (see, e.g., \cite{ibanez2007instanton}). Under these conditions, one can analytically derive the requirements for the generated neutrino masses and mixing angles to be consistent with experimental data.

\subsubsection{Normal hierarchy and neutrino mixing}
To begin, we assume a normal hierarchy for the neutrino masses, i.e. $m_1 \ll m_2 \approx m_3$, and set $m_1$ to zero. Using the expression in \eqref{UPMNS3}, the neutrino mass matrix takes the form
\begin{equation} \label{ML1}
M^L= U_{PMNS} \cdot \text{diag}(0, m_2 e^{i\alpha_2}, m_3) \cdot U^T_{PMNS}
\end{equation}
Comparing this structure with \eqref{MLij}, one possible way to reproduce it is to identify $I_2 = m_2$, $I_3 = m_3$ and to choose the coefficients $c^{(2)}_i$ and $c^{(3)}_i$ as
\begin{equation} \label{c2i1}
(c^{(2)}_1, c^{(2)}_2, c^{(2)}_3)= e^{i\alpha_2/2} (U_{e2}, U_{\mu 2}, U_{\tau 2}),
\end{equation}
\begin{equation} \label{c3i1}
(c^{(3)}_1, c^{(3)}_2, c^{(3)}_3)= (U_{e3}, U_{\mu 3}, U_{\tau 3}).
\end{equation}
It should be emphasized that this represents a specific choice rather than the most general scenario. Nevertheless, given the hierarchies among both the neutrino masses and the instanton contributions, it is natural to associate one instanton with the generation of $m_2$ and another with $m_3$. Note also that the normalization of the flavor vectors $c^{(r)}_i$ can be adjusted via $c^{(r)} \rightarrow \mathcal{N} c^{(r)}$ if one replaces $I_r =m_r$ with $I_r= \mathcal{N}^{-2} m_r$. 

\subsubsection{The inverse hierarchy case}
Experimentally, two possible orderings of the neutrino masses are allowed: the so-called normal ordering, where $m_1 < m_2 < m_3$, and the so-called inverse ordering, where $m_3 < m_1 < m_2$. In the latter case, if $m_3 \ll m_1 \approx m_2$, the neutrino spectrum is referred to as inversely hierarchical. String theory instantons can in principle also realize this scenario.
However, one must bear in mind that the splitting between $m_1$ and $m_2$ is very small, and an explanation for why $m_1 \approx m_2$ would be required. From the perspective of string instantons, this would necessitate the presence of two instantons, $D2_1$, $D2_2$ with approximately equal actions $S(1)$, $S(2)$, but with significantly different flavor vectors $c^{(1)}_a$, $c^{(2)}_a$. Since the action is essentially determined by the size of the wrapped 3-cycle in the extra dimensions, and these sizes are generally expected to differ, a certain degree of fine-tuning would be needed. Different values for the various actions $S^{(r)}$, typically leading to some hierarchy, seem more generic. For completeness, however, we will also consider the inverted hierarchy case.

Following the strategy outlined in the previous subsection, examples of patterns for the relevant coefficients $c^{(1)}_i$ and $c^{(2)}_i$ in this scenario can be readily obtained. Setting $m_3=0$, the neutrino mass matrix has the form
\begin{equation} \label{ML2}
M^L= U_{PMNS} \cdot \text{diag}(m_1 e^{i\alpha_1}, m_2 e^{i\alpha_2}, 0) \cdot U^T_{PMNS}
\end{equation}
and only two string instantons are required (in the minimal case) to generate the neutrino masses $m_1$ and $m_2$. Again, comparing with \eqref{MLij}, we find that one possible choice is $I_1 = m_1$, $I_2 = m_2$ (with $I_1 \approx I_2$), along with
\begin{equation} \label{c1i2}
(c^{(1)}_1, c^{(1)}_2, c^{(1)}_3)= e^{i\alpha_1/2} (U_{e1}, U_{\mu 1}, U_{\tau 1}),
\end{equation}
\begin{equation} \label{c2i2}
(c^{(2)}_1, c^{(2)}_2, c^{(2)}_3)= e^{i\alpha_2/2} (U_{e2}, U_{\mu 2}, U_{\tau 2}).
\end{equation}

\subsubsection{Quasi-degenerate neutrino spectrum}
We now turn to the general case, which includes scenarios of quasi-degenerate (or partially degenerate) neutrino masses, where $m_1$, $m_2$, and $m_3$ are all non-zero and typically two of them are nearly degenerate. In principle, our scheme can also accommodate this situation. However, in this case the splittings between $m_1$, $m_2$, and $m_3$ are very small, and this near degeneracy of the mass eigenvalues would need to be explained. Specifically, the masses must satisfy the experimental constraints: i.e. $m^2_2 - m^2_1 \approx 7.9 \times 10^{-5} \text{eV}^2$, $|m^2_3 - m^2_1| \approx 2.6 \times 10^{-3} \text{eV}^2$, while $m_1 \approx m_2 \approx m_3$ are much larger than the mass splittings. From the perspective of string instantons, this would again require three different instantons with nearly identical actions but very different flavor vectors. Although such a configuration is possible, it would involve a certain amount of fine-tuning and is not generically expected.

To provide an example of a pattern for the coefficients $c^{(r)}_i$ that is compatible with the experimentally observed mixing angles, quasi-degenerate neutrinos could be realized using three instantons with $c^{(1)}$, $c^{(2)}$, $c^{(3)}$ chosen as
\begin{equation} \label{c1i3}
(c^{(1)}_1, c^{(1)}_2, c^{(1)}_3)= e^{i\alpha_1/2} (U_{e1}, U_{\mu 1}, U_{\tau 1})
\end{equation}
\begin{equation} \label{c2i3}
(c^{(2)}_1, c^{(2)}_2, c^{(2)}_3)= e^{i\alpha_2/2} (U_{e2}, U_{\mu 2}, U_{\tau 2})
\end{equation}
\begin{equation} \label{c3i3}
(c^{(3)}_1, c^{(3)}_2, c^{(3)}_3)= (U_{e3}, U_{\mu 3}, U_{\tau 3}).
\end{equation}
together with $I_1 = m_1$, $I_2 = m_2$, $I_3 = m_3$ (and consequently $I_1 \approx I_2 \approx I_3$).

\subsection{The see-saw case}
So far, we have considered the scenario in which the dominant contribution to the observed neutrino masses arises from the Weinberg operator. Let us now examine the opposite case, where the seesaw mechanism provides the leading contribution to neutrino masses. Deriving analytic formulas for the most general seesaw scenario is challenging. The seesaw contribution to the neutrino mass matrix can depend significantly on the structure of the neutrino Yukawa matrix $h_D$, giving rise to a wide variety of possible patterns for $h_D$ and $M^R_{ab}$ that are consistent with experimental neutrino data (again assuming small charged-lepton mixing, as before). In the remainder of Section 3.2, we discuss the special case in which only small mixing originates from the neutrino Yukawa matrix $h_D$. Specifically, we consider the limiting case where $h_D$ is diagonal, 
\begin{equation} \label{hD}
h_D= \text{diag} (y^{(\nu)}_e, y^{(\nu)}_{\mu}, y^{(\nu)}_{\tau} ). 
\end{equation}
Small mixing induced by $h_D$ can be treated as a perturbation and incorporated in a straightforward manner. We further assume that the seesaw contribution dominates the neutrino mass matrix.

The neutrino mass matrix is then given by
\begin{equation} \label{MLab}
M^L_{ab} = \left( \sum_r \frac{d^{(r)}_a}{(h^T_D)_{aa}} \frac{d^{(r)}_b}{(h^T_D)_{bb}} \tilde{I}^{-1}_r \right)^{-1},
\end{equation}
where
\begin{equation} \label{tildeIr}
\tilde{I}_r= \frac{\langle \bar{H} \rangle^2}{2 M_s} \frac{1}{e^{-S_r}}
\end{equation}
which in turn defines the quantity
\begin{equation} \label{tildeMLab}
\tilde{M}^L_{ab} \equiv (M^L_{ab})^{-1} =  \sum_r \frac{d^{(r)}_a}{(h^T_D)_{aa}} \frac{d^{(r)}_b}{(h^T_D)_{bb}} \tilde{I}^{-1}_r.
\end{equation}
The indices of the matrix $\tilde{M}^L_{ab}$ must be understood as arising from the numbers $d^{(r)}_a$ and $d^{(r)}_b$, whereas those of the matrix $M^L_{ab}$ are computed according to the usual rules of matrix calculus.

We now discuss how to choose the coefficients $d^{(r)}_a$ in order to realize neutrino mixing. The form of $\tilde{M}^L$ required to realize neutrino mixing is thus given by 
\begin{equation} \label{tML}
\begin{split}
\tilde{M}^L = & U \text{diag} (\frac{1}{m_1 e^{i\alpha_1}}, \frac{1}{m_2 e^{i\alpha_2}}, \frac{1}{m_3}) U^T \\
= & \frac{1}{m_1 e^{i\alpha_1}} 
\left[ \begin{array}{ccc}		
U^2_{e1} & U_{e1} U_{\mu1} & U_{e1} U_{\tau1} \\		
U_{e1} U_{\mu1} & U^2_{\mu1} & U_{\mu1} U_{\tau1} \\		
U_{e1} U_{\tau1} & U_{\mu1} U_{\tau1} & U^2_{\tau1}\\		
\end{array}	\right] + \frac{1}{m_2 e^{i\alpha_2}} 
\left[ \begin{array}{ccc}		
U^2_{e2} & U_{e2} U_{\mu2} & U_{e2} U_{\tau2} \\		
U_{e2} U_{\mu2} & U^2_{\mu2} & U_{\mu2} U_{\tau2} \\		
U_{e2} U_{\tau2} & U_{\mu2} U_{\tau2} & U^2_{\tau2}\\		
\end{array}	\right] \\
& + \frac{1}{m_3} 
\left[ \begin{array}{ccc}		
U^2_{e3} & U_{e3} U_{\mu3} & U_{e3} U_{\tau3} \\		
U_{e3} U_{\mu3} & U^2_{\mu3} & U_{\mu3} U_{\tau3} \\		
U_{e3} U_{\tau3} & U_{\mu3} U_{\tau3} & U^2_{\tau3}\\		
\end{array}	\right].
\end{split}
\end{equation}
A possible choice for the coefficients $d^{(r)}_a$ is therefore to take $\tilde{I}_1=m_1$, $\tilde{I}_2=m_2$, $\tilde{I}_3=m_3$ and 
\begin{equation} \label{d(1)123}
\left( \frac{d^{(1)}_1}{y^{(\nu)}_e}, \frac{d^{(1)}_2}{y^{(\nu)}_{\nu}}, \frac{d^{(1)}_3}{y^{(\nu)}_{\tau}} \right) = e^{-i \alpha_1/2} \left( U_{e1}, U_{\mu 1}, U_{\tau1} \right),
\end{equation}
\begin{equation} \label{d(2)123}
\left( \frac{d^{(2)}_1}{y^{(\nu)}_e}, \frac{d^{(2)}_2}{y^{(\nu)}_{\nu}}, \frac{d^{(2)}_3}{y^{(\nu)}_{\tau}} \right) = e^{-i \alpha_2/2} \left( U_{e2}, U_{\mu 2}, U_{\tau2} \right),
\end{equation}
\begin{equation} \label{d(3)123}
\left( \frac{d^{(3)}_1}{y^{(\nu)}_e}, \frac{d^{(3)}_2}{y^{(\nu)}_{\nu}}, \frac{d^{(3)}_3}{y^{(\nu)}_{\tau}} \right) = \left( U_{e3}, U_{\mu3}, U_{\tau3} \right).
\end{equation}
We observe that if $(h_D)_{aa} = y^{(\nu)}_a$, $a = e, \mu, \tau$ exhibit a hierarchical structure, then the coefficients $d^{(r)}_a$ (for all $r$) would also need to follow a very similar hierarchical pattern in order to generate large neutrino mixing.

More generally, instantons can generate the Weinberg operator responsible for neutrino masses. After electroweak symmetry breaking, this operator yields a direct mass term for some (or all) of the three light neutrinos, as well as the Majorana mass matrix for the right-handed neutrinos.

\subsection{The mass ordering of left-handed neutrinos}
In this subsection, we determine the mass ordering of left-handed Majorana neutrinos. We adopt the following parameters: $\langle \bar{H} \rangle =246 \text{GeV}$ and $M_s = 10^{17} \text{GeV}$. This yields 
\begin{equation} \label{2H2Ms}
\frac{2 \langle \bar{H} \rangle^2}{M_s} = 1.2103 \times 10^{-12} \text{GeV} =1.2103 \times 10^{-3} \text{eV},
\end{equation}
\begin{equation} \label{H22Ms}
\frac{\langle \bar{H} \rangle^2}{2 M_s} = 3.0258 \times 10^{-13} \text{GeV} =3.0258 \times 10^{-4} \text{eV}.
\end{equation}
First, we consider the case where the Weinberg operator dominates. Using $\Delta m^2_{21} = m^2_2 - m^2_1 = 7.9 \times 10^{-5} \text{eV}^2$, we obtain
\begin{equation} \label{Deltam221}
\Delta m^2_{21} = m^2_2 - m^2_1 = \left( \frac{2 \langle \bar{H} \rangle^2}{M_s} \right)^2 \left( e^{-2U_2} -e^{-2U_1} \right) = 7.9 \times 10^{-5} \text{eV}^2>0,
\end{equation}
which implies $e^{-2U_2} -e^{-2U_1} = 53.9323$. Since $U_r$ describes a shape parameter of the extra dimensions and is typically related to a geometric scale, it should naturally be positive. Therefore, this result is unreasonable. 

Next, we consider the case where the seesaw mechanism dominates. Again using $\Delta m^2_{21} = m^2_2 - m^2_1 = 7.9 \times 10^{-5} \text{eV}^2$, we have
\begin{equation} \label{Deltam221}
\Delta m^2_{21} = m^2_2 - m^2_1 = \left( \frac{ \langle \bar{H} \rangle^2}{2 M_s} \right)^2 \left( e^{2U_2} -e^{2U_1} \right) = 7.9 \times 10^{-5} \text{eV}^2,
\end{equation}
giving $e^{2U_2} -e^{2 U_1} = 862.87$, which is reasonable. If the neutrino mass ordering is normal $(m_1<m_2<m_3)$, then using $\Delta m^2_{31} = m^2_3 - m^2_1 = 2.6 \times 10^{-3} \text{eV}^2$, we obtain
\begin{equation} \label{Deltam231}
\Delta m^2_{31} = m^2_3 - m^2_1 = \left( \frac{ \langle \bar{H} \rangle^2}{2 M_s} \right)^2 \left( e^{2U_3} -e^{2U_1} \right) = 2.6 \times 10^{-3} \text{eV}^2>0,
\end{equation}
which leads to $e^{2U_3} -e^{2U_1} = 28398.23$. We can find solutions that satisfy the sum of the masses of left-handed neutrinos ($m_{tot}$) of the observed data \cite{navas2024review,planck2020planck}.

On the contrary, if the neutrino mass ordering is inverse $(m_3<m_1<m_2)$, we have $e^{2U_3} -e^{2U_1} = -28398.23$. The critical solution is $U_1=5.12702$ and $U_3=0$. So the critical mass of $m_1$ and $m_3$ are 
\begin{equation} \label{iom1cri}
m^{cri}_1=3.0258 \times 10^{-4} \times e^{5.12702} \text{eV}= 0.05099\text{eV} 
\end{equation}
and 
\begin{equation} \label{iom3cri}
m^{cri}_3=3.0258 \times 10^{-4} \times e^{0} \text{eV}= 3.0258 \times 10^{-4} \text{eV}. 
\end{equation}
The corresponding $m^{cri}_2$ is then given by
\begin{equation} \label{iom2cri}
m^{cri}_2=\sqrt{(m^{cri}_2)^2 + \Delta m^2_{21}} =0.05176\text{eV}. 
\end{equation}
Therefore, the total mass of three left-handed neutrino is $m_{tot} = m^{cri}_1 + m^{cri}_2 + m^{cri}_3 =0.10305 \text{eV}$, which is larger than the observational data of $m_{tot} \leq 0.06 \sim 0.07\text{eV}$ \cite{navas2024review,planck2020planck}. Therefore, the inverse ordering is unreasonable.

Therefore, we conclude that left-handed Majorana neutrino masses originate predominantly from the seesaw mechanism, and the left-handed neutrino mass hierarchy should be normal, i.e. $m_1 < m_2 < m_3$. This result can be tested in next-generation experiments.

\section{Phenomenological comparison with other neutrino mass mechanisms}
The D2-instanton induced Majorana neutrino mass mechanism in Type IIA intersecting D6-brane models exhibits several distinctive features when compared with the standard neutrino mass generation scenarios. In this section we contrast our framework with the type II seesaw, radiative neutrino mass models, Dirac neutrino scenarios, and the inverse seesaw, and further compare it with alternative string-theoretic approaches such as heterotic orbifolds and F-theory GUTs.

\subsection{Comparison with field-theoretic mechanisms}
\subsubsection{Type II seesaw}
In the canonical type II seesaw, a scalar $SU(2)_L$ triplet $\Delta$ is introduced, and the neutrino mass matrix is given by $M_{\nu}=f\langle \Delta^0 \rangle$, where $\langle \Delta^0 \rangle \approx \mu v^2/M^2_{\Delta}$ is induced through the trilinear coupling $\mu \Phi^T i \tau_2 \Delta \Phi$ \cite{mandal2022toward}. The smallness of neutrino masses thus relies on a parametrically small induced VEV, which in turn requires either a small $\mu$ term or a large triplet mass $M_{\Delta}$. Both regimes demand an additional layer of explanation -- either a finely tuned trilinear coupling or a large mass scale. In contrast, the D2-instanton mechanism generates the Majorana mass directly through the exponentially suppressed factor $e^{-U_{\text{inst}}}$, where the exponent is the volume of a wrapped 3-cycle and is not tied to any Standard Model gauge coupling. Consequently, the suppression is “geometrically natural" rather than requiring an artificially small parameter.

\subsubsection{Radiative neutrino mass models}
In radiative frameworks \cite{cai2017trees}, neutrinos are massless at tree level and acquire mass at one or more loops, with the suppression coming from loop factors $\sim (16 \pi^2)^{-n}$ and the breaking of lepton number by a small coupling. The new particles in these models are typically at the TeV scale, making them testable at colliders but requiring a specific symmetry structure to forbid the tree-level Weinberg operator. Our D2-instanton mechanism, by contrast, produces the dimension-5 Weinberg operator directly at the string scale \cite{cvetivc2007computation}:
\begin{equation} \label{d5Wo}
\mathcal{W}_{\text{W}}= e^{-U_{\text{inst}}} \frac{1}{M_s} (L^a \bar{H})(L^b \bar{H}),
\end{equation}
with the flavor structure factorized as $\bar{c}^{(r)}_a c^{(r)}_b$. The hierarchy among neutrino masses emerges from the different instanton actions $U_r$, which are generically different because they correspond to different wrapped volumes. This provides a natural explanation for the observed hierarchy without invoking loop suppression or additional symmetries.

\subsubsection{Dirac neutrino scenarios}
The simplest Dirac neutrino model requires Yukawa couplings $y_{\nu} \sim 10^{-12}$ \cite{behera2025phenomenology}, which is technically natural but phenomenologically unsatisfactory. More elegant Dirac seesaw models introduce additional vector-like fermions and a $\mathbb{Z}_2$ symmetry, with the light neutrino mass $m_{\nu} \sim m_L m_R/M_N$ arising from a hierarchy $m_L \ll m_R \ll M_N$. Even in these constructions, one must impose a global $U(1)_L$ lepton number symmetry to forbid the Majorana mass term. In our Type IIA framework, the Majorana nature is automatic: the D2-instanton carries $B-L$ charge $+2$ and generates the $\nu_R \nu_R$ operator directly, making neutrinos inherently Majorana particles. 

\subsubsection{Inverse seesaw}
The inverse seesaw \cite{behera2025phenomenology} introduces two sets of fermionic singlets $N_R$ and $S_L$, with the light neutrino mass $m_{\nu} \approx m_D M^{-1} \mu(M^T)^{-1} m^T_D$ suppressed by a small lepton-number-violating parameter $\mu$. The smallness of $\mu$ is protected by an approximate $U(1)_L$ symmetry which is only softly broken. While technically natural, this mechanism requires postulating a small $\mu$ and a specific symmetry structure. In the D2-instanton mechanism, the smallness of neutrino masses is geometric -- it arises from the exponential suppression $e^{-\text{Re} U_r}$, where $\text{Re} U_r$ is the volume of the wrapped cycle. No additional symmetry or small parameter needs to be imposed; the hierarchy is a generic outcome of the string compactification.

\subsection{The hierarchy of right-handed neutrino masses}
A key advantage of the D2-instanton mechanism is its natural generation of mass hierarchy. The right-handed Majorana neutrino mass matrix takes the factorized form:
\begin{equation} \label{rhnM}
M^R_{ab}= 2 M_s \sum_r d^{(r)}_a d^{(r)}_b e^{-\text{Re} U_r}.
\end{equation}
Since the exponents $\text{Re} U_r$ are volumes of distinct 3-cycles, they are generically different. A modest hierarchy among the  $\text{Re} U_r$ values -- perfectly natural in geometric compactifications -- directly translates into a hierarchy among the eigenvalues of $M^R$ and, via the seesaw formula, among the light neutrino masses. In our numerical example (Section 6), we use $e^{-\text{Re}U_1}=0.125$, $e^{-\text{Re}U_2}=0.0325$, $e^{-\text{Re}U_3}=0.0059$, yielding a normal ordering $m_1 < m_2 < m_3$. This stands in sharp contrast to:
\begin{itemize}
 \item Type II seesaw, where the hierarchy must be put in by hand through the structure of the Yukawa matrix $f_{\alpha \beta}$.
 \item Inverse seesaw, where the smallness of $\mu$ must be protected by an approximate symmetry.
 \item Radiative models, where the hierarchy arises from loop suppression but requires specific symmetry assignments to forbid tree-level masses.
 \item Dirac models, where the hierarchy (if any) comes from the imposed $U(1)_L$ charge structure.
\end{itemize}
Relative to these alternatives, the D2-instanton framework has the virtue that the suppression is geometric in origin and not tied to a small symmetry-breaking parameter. We stress, however, that establishing the hierarchy as a prediction requires computing $U_r$ and $d^{(r)}_a$ in a specific compactification, which we do not attempt here.

\subsection{Prediction for $\langle m_{\beta \beta} \rangle$}
The effective Majorana mass probed in neutrinoless double beta decay is \cite{vergados2016neutrinoless}
\begin{equation} \label{ntdbd}
\langle m_{\beta \beta} \rangle = |\sum^3_{i=1} U^2_{ei} m_i|.
\end{equation}
For normal ordering, the expected range is $\langle m_{\beta \beta} \rangle \sim 1-5\text{meV}$. Our numerical example yields $\langle m_{\beta \beta} \rangle = 2.594 \text{meV}$, squarely within this window and accessible to next-generation experiments. This prediction is a generic consequence of our framework: the D2-instanton mechanism naturally produces normal ordering (as argued in Section 3.3) and yields $\langle m_{\beta \beta} \rangle$ in the 1–10 meV range. By contrast:
\begin{itemize} 
\item Inverse ordering (which some Type II seesaw and radiative models can accommodate) would predict $\langle m_{\beta \beta} \rangle \sim 15-45\text{meV}$.
\item Quasi-degenerate spectra (possible in some inverse seesaw constructions) would give $\langle m_{\beta \beta} \rangle \sim 50-200\text{meV}$.
\end{itemize}
Thus, a measurement of $\langle m_{\beta \beta} \rangle$ in the $1–10\text{meV}$ range would strongly favor our geometric hierarchy mechanism over scenarios predicting inverse ordering or quasi-degeneracy.

\subsection{Comparison with other string-theoretic frameworks}
\subsubsection{Heterotic orbifolds}
In heterotic orbifold constructions \cite{giedt2005massive}, a detailed analysis of the $\mathbb{Z}_3$ orbifold revealed that the minimal seesaw mechanism does not arise naturally -- the required simultaneous presence of Dirac and Majorana mass terms for right-handed neutrinos is obstructed by string selection rules. Alternative mechanisms such as extended seesaw, small Dirac masses, or triplet seesaw must be invoked. This highlights a significant challenge: heterotic models often struggle to generate the right combination of operators. In contrast, Type IIA intersecting D6-brane models explicitly allow D2-instantons to generate both the $\nu_R \nu_R$ Majorana mass and the Weinberg operator, as demonstrated in concrete local $SU(5)$-like GUT setups on $T^6/(\mathbb{Z}_2 \times \mathbb{Z}'_2)$ with discrete torsion. The non-perturbative origin is genuinely stringy and not available in heterotic perturbation theory.

\subsubsection{F-theory GUTs}
F-theory GUT models generate neutrino masses through the seesaw mechanism \cite{tatar2009right,Bouchard:2009bu}, with the hierarchy arising from the dilution of wave function overlaps and Kaluza-Klein mode effects. The predicted mass ratios are roughly $m_1:m_2:m_3 \sim \alpha_{\text{GUT}}:\alpha^{1/2}_{\text{GUT}}:1$, which is in reasonable accord with observations. However, F-theory predictions for $\theta_{13}$ tend to be large ($\theta_{13}\sim \theta_{C}\sim 0.2$), which is at the upper edge of the current experimental range. Moreover, the precise structure of the Yukawa matrices in F-theory depends sensitively on the geometry of the GUT seven-brane and the localization of wave functions. In our Type IIA framework, the flavor structure is encoded in the instanton zero-mode couplings $d^{(r)}_a$ and $c^{(r)}_a$, which are computable from worldsheet correlators and Jacobi theta functions. This provides a more direct handle on flavor predictions.

\subsubsection{The advantage of Type IIA intersecting brane}
The key advantage of the Type IIA route is twofold:
\begin{itemize}
 \item Non-perturbative origin: The Majorana mass and Weinberg operator arise from D2-instantons, whose exponential suppression is not related to any SM gauge coupling. This avoids the severe fine-tuning that would be required if the suppression came from a gauge instanton.
 \item Geometric origin of the suppression: The exponential factor $e^{-\text{Re}U_r}$ is a wrapped three-cycle volume, not a gauge coupling, so the suppression is not tied to any small Standard Model parameter. Different instantons wrapping different cycles automatically produce different suppression factors, and this is a genuine structural advantage over mechanisms requiring an artificially small coupling. We caution, however, that the light neutrino hierarchy also depends on the flavour vectors, which in the present work are not computed (see Section 4.2 and Section 6.3).
\end{itemize}
This is in contrast to heterotic models where the seesaw structure is often obstructed, or F-theory models where the hierarchy relies on subtle wave function overlaps.

To summarize, the D2-instanton mechanism in Type IIA intersecting D6-brane models is distinguished by natural hierarchy, automatic Majorana nature, predictive $\langle m_{\beta \beta} \rangle$, string-theoretic origin and compatibility with leptogenesis. These features collectively position the Type IIA D2-instanton framework as a highly attractive and testable string-theoretic solution to the neutrino mass and baryon asymmetry puzzles.

\section{The constraints of swampland conjecture}
In the $\mathcal{N}=1$ Type IIA orientifold $T^6/(\mathbb{Z}_2 \times \mathbb{Z}_2)$ with intersecting D6‑branes, the MSSM chiral spectrum arises from intersection numbers $I_{ab} = [\Pi_a] \cdot [\Pi_b]$ of the wrapped 3‑cycles, while Yukawa couplings originate from triple‑intersection points. Baryon-number violation and lepton-number violation $(\Delta B, \Delta L)$ are not generated at the perturbative level; it must come from non‑perturbative Euclidean D2 (E2) instantons. Such E2 instantons can generate a right-handed neutrino Majorana mass term that is otherwise forbidden; when combined with the perturbative Dirac mass, this enables the seesaw mechanism \cite{cvetivc2007computation}. In a globally consistent intersecting D6 model, CP violation can arise both from geometric phases in the Yukawa couplings and from non‑perturbative phases of the E2 instantons, while the observed baryon asymmetry is generated through instanton‑mediated operators combined with moduli driven leptogenesis \cite{Liu:2025zjs}.

\subsection{Swampland Distance Conjecture (SDC) bounds the instanton action}
The SDC asserts that as one goes to infinite distance in moduli space a tower of light states appears, signaling the breakdown of the $4D$ effective field theory \cite{palti2019swampland}. In the D6 context the relevant “distance” is the volume of the 3-cycle wrapped by the E2 instanton. The instanton action is \cite{cvetivc2007computation}
\begin{equation} \label{instanaction}
S_{E2} = \frac{\text{Vol(E2)}}{2 \pi \alpha'}, \quad c_{\slashed{L}} \sim e^{-S_{E2}}.
\end{equation}
If one tried to make $\Delta L=2$ arbitrarily small (or large) by sending $\text{Vol}(E2) \rightarrow \infty$, the $m_{KK}$ is driven to zero as the internal volume is increased, destroying the validity of the $4D$ description. Hence $S_{E2}$ is bounded from below, and the resulting Weinberg operator scale
\begin{equation} \label{Wos}
\frac{1}{\Lambda_W} \sim \frac{e^{-S_{E2}}}{M_s}
\end{equation}
is forced into a window that typically yields $m_{\nu} \sim \text{meV}-0.1\text{eV}$, compatible with observation but precluding any fine‑tuned dialing of the neutrino mass. This geometric lock is the D6‑specific realization of the SDC.

The Swampland Distance Conjecture (SDC) implies that the modulus field controlling the instanton action cannot be taken to infinity without a tower of light states appearing, which would break the effective field theory. In Type IIA intersecting D6-brane models, the relevant modulus is the Kähler modulus that determines the volume of the 3-cycle wrapped by the E2-instanton. A standard application of the SDC to such setups gives a lower bound on the instanton action \cite{ooguri2007geometry}:
\begin{equation} \label{lbSE2a}
S_{E2} \equiv \text{Re} U \geq \ln \left( \frac{M_{P}}{M_s} \right) \approx \ln \left( \frac{1.22 \times 10^{19} \text{GeV}}{10^{17}\text{GeV}}\right) = 4.8
\end{equation}
This is a conservative estimate; in concrete compactifications the bound can be weaker due to $O(1)$ factors.

In our numerical example (Section 6), the three instanton actions are
\begin{equation} \label{REUr123}
\text{Re}U_1=2.079, \quad \text{Re}U_2=3.427, \quad \text{Re}U_3 = 5.133.
\end{equation}
We see that $\text{Re}U_3$ exceeds the nominal bound, while $\text{Re}U_1$ and $\text{Re}U_2$ are smaller. However, the SDC bound is not a strict cutoff but rather a parametric estimate; in explicit Type IIA orientifold models with moduli stabilization, the actual allowed range can be shifted by factors of order unity \cite{cvetivc2007computation}. Moreover, the modulus excursion associated with these instanton actions may be smaller than the naive field range, relaxing the bound. Therefore, the values $\text{Re}U_1=2.079$ and $\text{Re}U_2=3.427$ are not ruled out; they lie in a marginal region that requires a full moduli stabilization analysis to decide definitively. For the purpose of this phenomenological study, we take them as acceptable, noting that the largest instanton action $\text{Re}U_3$ is safely above the bound.

\subsection{dS conjectures suppress the reheating temperature}
The refined dS conjecture (and its stronger version, the Trans‑Planckian Censorship Conjecture) imposes a lower bound on $|\nabla V|$ or restricts the duration of inflation, which in turn caps the Hubble scale during inflation and the subsequent reheating temperature \cite{palti2019swampland}. In the concrete setting of classical Type IIA flux compactifications with (anti‑)O6‑planes and (anti‑)D6‑branes, parametrically controlled dS vacua are strongly constrained; at parametrically large volume or parametrically small string coupling such dS vacua are ruled out \cite{junghans2019weakly}. The chain of constraints is:
\begin{equation} \label{dScontem}
\text{dS conjecture}/\text{TCC} \rightarrow H_{\text{inf}} \, \, \text{small} \rightarrow T_{\text{RH}} \leq 10^{6} \text{GeV}-10^9 \text{GeV}. 
\end{equation}
Therefore any baryogenesis mechanism that requires $T_{\text{RH}}$ above this upper bound is excluded. This eliminates:
\begin{itemize} 
\item GUT scale baryogenesis ($T \sim M_X \sim 10^{15-16}\text{GeV}$) -- impossible.
\item Hierarchical thermal leptogenesis $(M_{N1}) \geq 5 \times 10^8 \text{GeV}$ -- the Davidson-Ibarra bound demands $T_{\text{RH}} \geq M_{N1}$; with the swampland-forced low $T_{\text{RH}}$ the heavy neutrinos are never thermally produced.
\item Departure from thermal equilibrium -- the rolling of moduli (Kähler, complex‑structure, D-brane positions) after inflation provides a time-dependent background, allowing a non‑thermal generation of the lepton asymmetry.
\end{itemize}
What survives are non‑thermal or moduli driven channels, precisely the ones realized in the intersecting D6 framework.

If the bound of refined dS conjecture and Trans-Planckian Censorship Conjecture holds, then the thermal production of $N_1$ with $M_1 =3.1 \times 10^{10}\text{GeV}$ (we take in Section 6) would be impossible, and the baryon asymmetry must arise from a non-thermal mechanism, such as moduli driven leptogenesis. However, swampland conjecture is still a controversial research area. The validity of these conjectures remains unknown. In the present work, we do not consider the constraints from swampland conjecture. We leave an analysis of the stricter TCC scenario for future work. This shows that the viability of thermal leptogenesis in this paper depends on which swampland criterion is applied -- a nontrivial connection between neutrino physics and quantum gravity constraints.

\section{A numerical example: neutrino physics and matter-antimatter asymmetry}
In this section, we present a detailed numerical benchmark scenario that demonstrates how the Type IIA string instanton framework can simultaneously reproduce the experimental neutrino oscillation data and address the observed baryon asymmetry of the Universe. We proceed step-by-step from low-energy neutrino observables and stringy input parameters to the final prediction for the baryon-to-photon ratio, as well as the effective Majorana mass relevant for neutrinoless double-beta decay searches.

It is important to be explicit about the direction of the logic. The observable quantities computed here -- the light neutrino masses, the mixing angles, and the baryon asymmetry -- are determined by the inputs listed in Section 6.1 together with the Casas–Ibarra matrix $R$; they do not depend on the string scale or on the instanton actions. The stringy quantities enter only through Eq. \eqref{rhMmm7}, which determines the flavour vectors $d_r$ once $M_r$, $M_s$ and $U_r$ are specified. In this sense Section 6 is a consistency check on the mechanism rather than a prediction of it: it establishes that a set of flavour vectors and instanton actions exists for which the Type IIA instanton framework reproduces the data, and it exhibits that set.

\subsection{Input parameters}
We classify the inputs as follows. Taken from data: the light neutrino masses $m_i$ and the mixing angles and phases. Chosen on phenomenological grounds: the right-handed neutrino masses $M_i$, placed in the window where thermal leptogenesis is viable; the string scale $M_s$; and the Casas–Ibarra matrix $R$. Determined subsequently: the flavour vectors $d_r$, via Eq. \eqref{rhMmm7}.

We first collect all input parameters for the benchmark scenario, which assumes a normal neutrino mass hierarchy as favored by current global oscillation fits.

We adopt the following light neutrino mass eigenvalues, consistent with the measured solar and atmospheric mass splittings:
\begin{equation} \label{mL11}
m_1 = 9.998224 \times 10^{-4} \text{eV}, 
\end{equation}
\begin{equation} \label{mL21}
m_2 = 8.6992929 \times 10^{-3} \text{eV}, 
\end{equation}
\begin{equation} \label{mL31}
m_3 = 0.05049009 \text{eV}.
\end{equation}
The corresponding mass squared differences are
\begin{equation} \label{deltam2L1}
m^2_{21} = m^2_2 -m^2_1= 7.4678 \times 10^{-5} \text{eV}^2, \quad m^2_{31} = m^2_3 -m^2_1= 2.5482 \times 10^{-3} \text{eV}^2
\end{equation}
in excellent agreement with experimental measurements. The sum of light neutrino masses is
\begin{equation} \label{mt1}
m_t = \sum^3_{i=1} m_i =0.0602 \text{eV},
\end{equation}
which is within the range of observed value from Planck CMB data \cite{navas2024review,planck2020planck}. 

The model features three dominant D2-instanton contributions to the right-handed neutrino Majorana masses. The real parts of the corresponding complex moduli $U_r$ determine the exponential suppression of each instanton amplitude, set by the volume of the wrapped internal 3-cycle. We take:
\begin{equation} \label{ReU11}
e^{-\text{Re}U_1}=0.125 \: \rightarrow \: \text{Re}U_1=2.079,
\end{equation}
\begin{equation} \label{ReU21}
e^{-\text{Re}U_2}=0.0325 \: \rightarrow \: \text{Re}U_2=3.427,
\end{equation}
\begin{equation} \label{ReU31}
e^{-\text{Re}U_3}=0.0059 \: \rightarrow \: \text{Re}U_3=5.133.
\end{equation}
These values represent moderately suppressed instanton actions, typical of intersecting D-brane compactifications where the instanton volume is not tied to the small Standard Model gauge couplings.

We use the following central values for the leptonic mixing angles and CP phases, consistent with global fits of neutrino oscillation data:
\begin{equation} \label{thetadeltaalpha1}
\theta_{12} =33.5^{\circ}, \, \theta_{23}=45^{\circ}, \, \theta_{13}= 8.5^{\circ}, \, \delta=234^{\circ}, \alpha_1 =0^{\circ}, \alpha_2=30^{\circ}.
\end{equation}
Here $\delta$ denotes the Dirac CP-violating phase, and $\alpha_{1,2}$ are the two Majorana CP phases.

We set the string scale to $M_s =10^{17} \text{GeV}$, characteristic of intermediate-scale string compactifications. The three right-handed neutrino mass eigenvalues are chosen to lie in the $10^{10}-10^{11} \text{GeV}$ range  \cite{navas2024review,planck2020planck}:
\begin{equation} \label{MR7}
M_1 = 3.0 \times 10^{10} \text{GeV}, \: M_2 = 7.5 \times 10^{10} \text{GeV}, \: M_3 = 1.0 \times 10^{11} \text{GeV}.
\end{equation}

\subsection{Construction of the PMNS mixing matrix}
The leptonic mixing matrix $U_{PMNS}$ is constructed in the standard PDG parameterization,
\begin{equation} \label{UPMNSU0PMaj}
U_{PMNS} = U_0 \cdot P_{\text{Maj}},
\end{equation}
where $U_0$ is the unitary mixing matrix containing the three mixing angles and the Dirac phase, and
\begin{equation} \label{PMaj7}
P_{\text{Maj}} = \text{diag} (e^{i \alpha_1/2}, e^{i\alpha_2/2},1)
\end{equation}
is the diagonal Majorana phase matrix.

Substituting the numerical values for the angles and phases, we obtain the full PMNS matrix:
\begin{equation} \label{UPMNS7}
U_{PMNS} =
\left[ \begin{array}{ccc}		
0.8247, & 0.5272+i0.1413, & -0.0869+i0.1196 \\		
-0.3390+i0.0705, & 0.5902+i0.2065, & 0.6993 \\		
0.4414+i0.0705, & -0.5489-i0.0987, & 0.6993\\		
\end{array}	\right].
\end{equation}
Throughout this work we assume that lepton mixing originates dominantly from the neutrino sector, with negligible contributions from the charged lepton sector. This is generically expected in string instanton models, where neutrino and charged lepton masses arise from distinct microscopic mechanisms.

\subsection{Right-handed neutrino mass matrix and flavor vectors}
As explained in Section 2, the right-handed Majorana mass matrix generated by D2-instantons with $\text{Sp}(2)$ Chan-Paton symmetry takes the factorized form \eqref{rhMmm2}. For simplicity, we adopt a diagonal flavor structure in this benchmark: each instanton couples dominantly to a single generation of right-handed neutrinos, so that $d^{(r)}_a = 0$ for $a \neq r$. In this case the right-handed neutrino mass matrix is diagonal, with eigenvalues
\begin{equation} \label{rhMmm7}  
M_r = 2 M_s (d_r)^2  e^{-\text{Re}U_r}.
\end{equation}
Solving for the flavor vectors gives:
\begin{equation} \label{d17}
d_1 = \sqrt{\frac{M_1}{2 M_s e^{-\text{Re}U_1}}} = 1.0954 \times 10^{-3},
\end{equation}
\begin{equation} \label{d27}
d_2 = \sqrt{\frac{M_2}{2 M_s e^{-\text{Re}U_2}}} = 3.397 \times 10^{-3},
\end{equation}
\begin{equation} \label{d37}
d_3 = \sqrt{\frac{M_3}{2 M_s e^{-\text{Re}U_3}}} = 9.206 \times 10^{-3}.
\end{equation}
The hierarchical structure of $d_r$ reflects the hierarchy of right-handed neutrino masses, as naturally expected from the volume dependence of instanton actions in compactified string theory.

\subsection{Neutrino Yukawa matrix}
We construct $Y_{\nu}$ via the Casas-Ibarra parameterization, which expresses the Yukawa matrix in terms of the light neutrino mass eigenvalues $m_{\nu}$, the right-handed neutrino masses, the PMNS matrix, and a complex orthogonal matrix -- the Casas–Ibarra matrix $R$ that parameterizes the unconstrained degrees of freedom in the seesaw mechanism. Concretely,
\begin{equation} \label{YnuCI}
Y_{\nu} = \frac{i}{v} \sqrt{M^R} R \sqrt{m_{\nu}} U^{\dagger}_{PMNS},
\end{equation}
where $\sqrt{M^R} =\text{diag} (\sqrt{M_1}, \sqrt{M_2}, \sqrt{M_3})$ and $\sqrt{m_{\nu}} =\text{diag} (\sqrt{m_1}, \sqrt{m_2}, \sqrt{m_3})$. 
For this benchmark we choose a specific rotation matrix $R$ to generate a sizable CP asymmetry for leptogenesis, while remaining consistent with low-energy neutrino data. More details of matrix $R$ can be found in Appendix A. The resulting Dirac Yukawa matrix reads:
\begin{equation} \label{Ynu7}
Y^{\nu} 
\approx 10^{-3} \times 
\left[ \begin{array}{ccc}		
0.873+i0.302, & -0.893+i3.348, & -0.753+i0.392 \\		
0.656+i2.447, & 0.959+i2.740, & -0.458-i2.548  \\		
1.194-i2.187, & 1.409+i7.011, & 1.663+i5.977\\		
\end{array}	\right].
\end{equation}
All entries are of order $10^{-3}$, comparable in size to charged-lepton Yukawa couplings. We note that this matrix is reconstructed via Casas–Ibarra rather than computed from the compactification, so its magnitude reflects the input spectrum $M_i$ and $m_i$; we do not claim it is derived from brane geometry, though the order of magnitude is consistent with what intersecting-brane models typically produce.

\subsection{Baryon asymmetry from thermal leptogenesis}
We now compute the baryon asymmetry generated via thermal leptogenesis, driven by the out-of-equilibrium decay of the lightest right-handed neutrino $N_1$ in the early Universe. The CP asymmetry parameter $\epsilon_1$ describes the relative difference between the decay rates of $N_1$ into leptons and antileptons. It is given by \eqref{e1Ynu2} and \eqref{fMjM1}. For the mass ratio $x = M^2_3/M^2_1 \approx 11.11$, we find $f(11.11) \approx 2.819$. Substituting all values yields $\epsilon_1 \approx 1.546 \times 10^{-6}$, which is a typical magnitude for successful thermal leptogenesis scenarios \cite{covi1996cp,giudice2004towards}.

The generated lepton asymmetry is partially erased by inverse decay and lepton-number-violating scattering processes, parameterized by the washout efficiency factor $\kappa_f$. The strength of washout is governed by the effective neutrino mass
\begin{equation} \label{enm}
\tilde{m}_1 \equiv \frac{(Y^{\dagger}_{\nu} Y_{\nu})_{11} v^2}{M_1} \approx 0.029 \text{eV}.
\end{equation}
The equilibrium mass scale is $m_{\star} \approx 1.1 \times 10^{-3} \text{eV}$. Therefore, the washout parameter is
\begin{equation} \label{K7}
K =\frac{\tilde{m}_1}{m_{\star}} \approx 26.36,
\end{equation}
corresponding to the strong washout regime. Using the standard analytical approximation for the efficiency factor in the strong washout limit, we have
\begin{equation} \label{kappa7}
\kappa_f \approx \frac{0.3}{K (\ln K)^{0.6}} \approx 5.589 \times 10^{-3}.
\end{equation}
The final baryon-to-photon ratio is then
\begin{equation} \label{fba27}  
\eta_B \approx \frac{28}{79} \frac{n^{eq}_N}{s} \epsilon_1 \kappa_f \approx 0.01 \epsilon_1 \kappa_f \approx 8.64 \times 10^{-11},
\end{equation}
within an order of magnitude of the observed value $\eta^{obs}_B = 6.1 \times 10^{-10}$ from Planck CMB measurements \cite{navas2024review,planck2020planck}.

\subsection{Effective Majorana mass for neutrinoless double-beta decay}
A key phenomenological signature of Majorana neutrinos is neutrinoless double-beta $(0\nu\beta\beta)$ decay, whose rate is proportional to the square of the effective Majorana mass
\begin{equation} \label{effneumass71}
\langle m_{\beta\beta} \rangle \equiv |\sum^3_{i=1} U^2_{ei} m_i|.
\end{equation}
Substituting our benchmark parameters, we obtain
\begin{equation} \label{effneumass72}
\langle m_{\beta\beta} \rangle \approx 2.594 \times 10^{-3} \text{eV} =2.594 \text{meV}.
\end{equation}
This value lies within the sensitivity reach of next-generation $(0\nu\beta\beta)$ decay experiments, consistent with the normal-ordering expectation, though beyond the reach of currently approved searches.

In summary, this numerical example demonstrates that the Type IIA string instanton model with intersecting D6-branes can simultaneously accommodate all current neutrino oscillation data and explain the observed matter-antimatter asymmetry of the Universe via thermal leptogenesis, while yielding a concrete prediction for neutrinoless double-beta decay.

\section{Parameter scan}
In this section, we scan the parameter space of flavor vector $d^{(r)}_a$ and right-handed nutrino masses. The scan varies only quantities that are neither fixed by observation nor derived from the compactification, so it probes the seesaw and leptogenesis sector. It does not probe the string sector, because the instanton actions and flavour vectors are not computed in this work. 

We scan six parameters with broad, uninformative priors: $M_1$ log-flat in $10^9\!-\!10^{12}$ GeV; $M_2/M_1$ flat in $[2,10]$; $M_3/M_2$ flat in $[1.5,5]$; $m_1$ flat in $[0,10^{-2}]$ eV, with $m_2$ and $m_3$ then following from the measured splittings; ${\rm Re}\,z$ flat in $[0,\pi]$; and ${\rm Im}\,z$ flat in $[-c,c]$ with $c=2,1,0.6$ to study prior sensitivity.

Drawing ${\rm Re}\,U_r$ from $[0.5,6]$ and $d_r$ log-uniformly from $[10^{-4},1]$, and requiring all three $M_r$ to lie in $[10^9,10^{12}]$ GeV with $M_1<M_2<M_3$, we find an acceptance of only $0.89\%$ over $10^6$ samples. The required flavour vectors cluster at $d_r\sim\{4\times10^{-4},\,2\times10^{-3},\,1\times10^{-2}\}$, with $16–84\%$ ranges of $[1.8\!-\!10.2]\times10^{-4}$, $[1.0\!-\!5.6]\times10^{-3}$ and $[4.0\!-\!23]\times10^{-3}$ respectively, as shown in Figure 1. Crucially, no sampled point with all $d_r$ of order unity falls in the window: the fraction with all $d_r\in[0.1,1]$ is $0.000$, and only $17\%$ have all $d_r\in[10^{-3},1]$.
\begin{figure}[h]
\centering
\includegraphics[width=0.85
\linewidth]{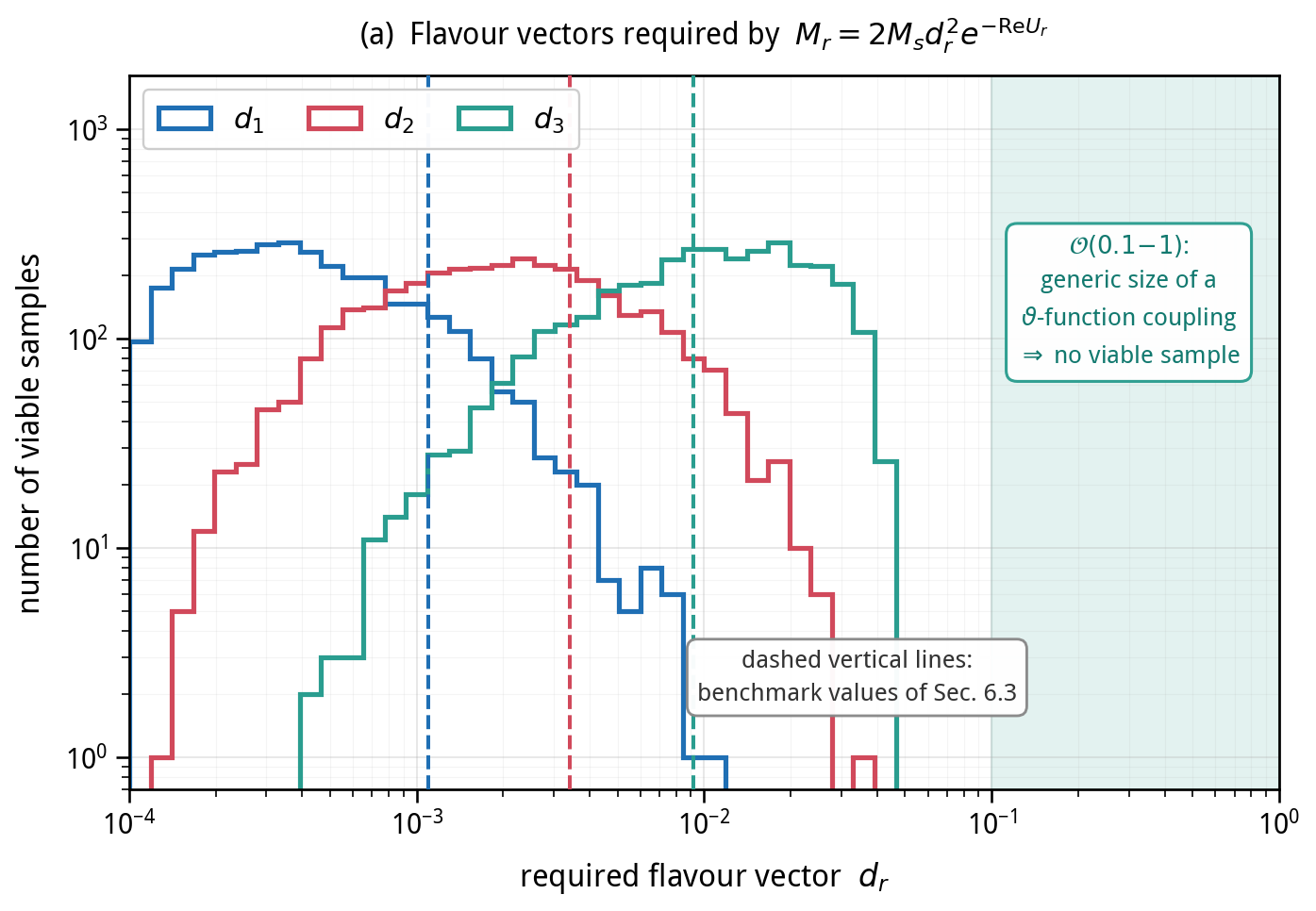}
\caption{Flavour vectors required by $M_r=2M_sd_r^2e^{-{\rm Re}U_r}$. Distributions of the three flavour vectors $d_r$ over the subset of samples for which all three right-handed masses fall in the leptogenesis window $10^9\!-\!10^{12}$ GeV with $M_1<M_2<M_3$, at $M_s=10^{17}$ GeV. Priors: ${\rm Re}\,U_r$ flat in $[0.5,6]$ and $d_r$ log-flat in $[10^{-4},1]$; $4\times10^5$ samples, of which $0.89\%$ are accepted. The three distributions peak at $d_r\simeq\{3.8\times10^{-4},\,2.0\times10^{-3},\,1.0\times10^{-2}\}$, with $16–84\%$ ranges $[1.8\!-\!10.2]\times10^{-4}$, $[1.0\!-\!5.6]\times10^{-3}$ and $[3.8\!-\!23]\times10^{-3}$. Dashed vertical lines mark the benchmark values of Sec. 6.3, $d_r=\{1.10,\,3.40,\,9.21\}\times10^{-3}$, which lie inside the accepted band. The green band indicates $\mathcal O(0.1\!-\!1)$, the generic magnitude of a $\vartheta$-function three-point coupling: no sampled point with all $d_r$ in this range falls in the window, the fraction being $0.000$.}
\label{Figure 1}
\end{figure}

Figure 2 makes the point directly. Setting $d_r=\mathcal O(1)$, the generic magnitude for a $\vartheta$-function three-point coupling, gives $M_r=\{2.5\times10^{16},\,6.5\times10^{15},\,1.2\times10^{15}\}$ GeV — too heavy by factors of $8.3\times10^5$, $8.7\times10^4$ and $1.2\times10^4$ relative to the required $\{3\times10^{10},7.5\times10^{10},10^{11}\}$ GeV. The benchmark of Section 6.3 requires instead $d_r=\{1.10,\,3.40,\,9.21\}\times10^{-3}$, which lies squarely inside the accepted band, so the benchmark is representative of what the mechanism needs rather than atypical. 
\begin{figure}[h]
\centering
\includegraphics[width=0.85
\linewidth]{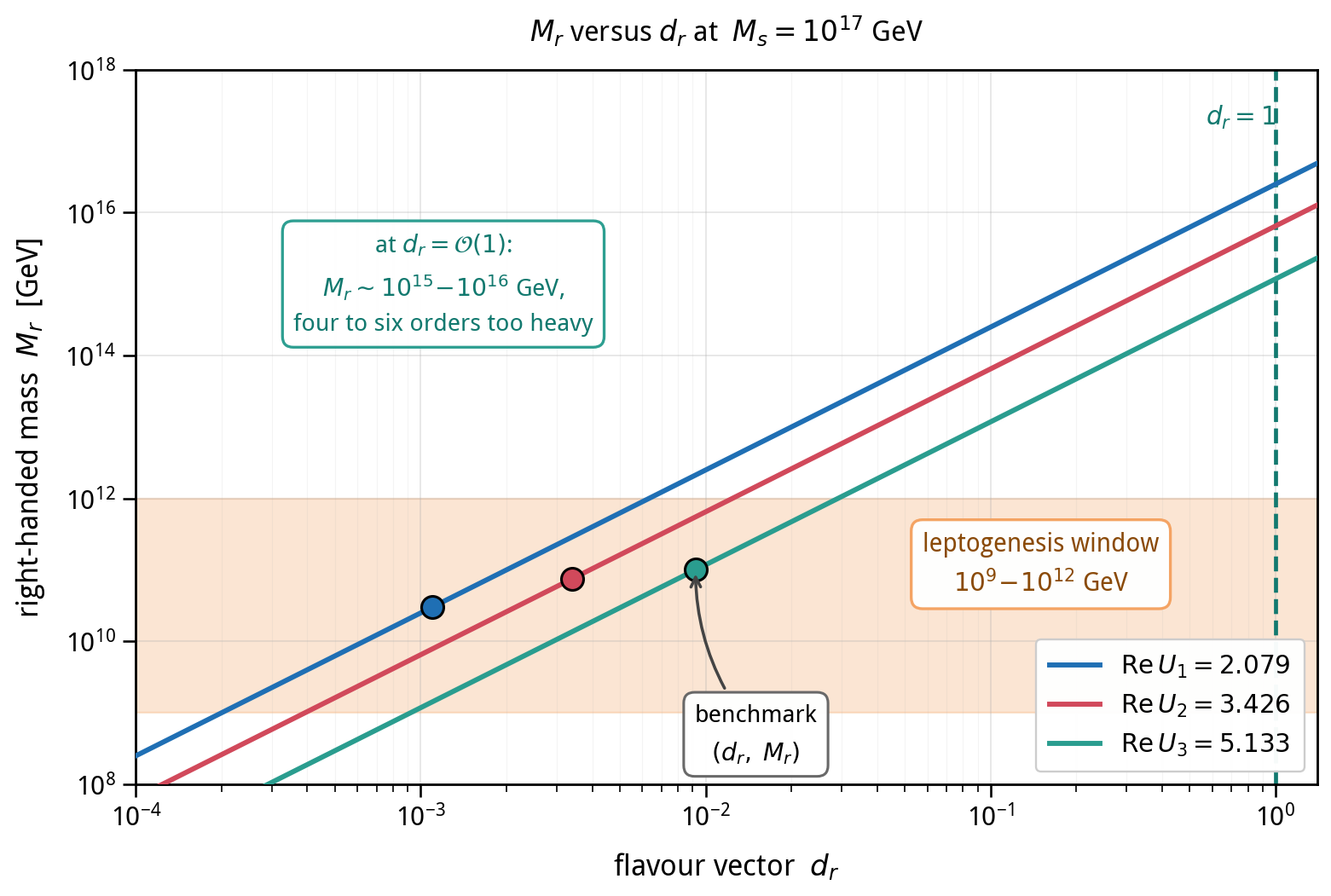}
\caption{Right-handed neutrino masses versus the flavour vectors at $M_s=10^{17}$ GeV. The three curves show $M_r=2M_s\,d_r^2\,e^{-{\rm Re}U_r}$ for the benchmark instanton actions ${\rm Re}\,U_r=\{2.079,\,3.427,\,5.133\}$. The orange band marks the window $10^9\!-\!10^{12}$ GeV in which thermal leptogenesis operates, and the filled circles mark the benchmark, $d_r=\{1.10,\,3.40,\,9.21\}\times10^{-3}$ paired with $M_r=\{3.0\times10^{10},\,7.5\times10^{10},\,1.0\times10^{11}\}$ GeV. The vertical dashed line marks $d_r=1$: at this value, generic for a worldsheet three-point coupling, the masses are $\{2.5\times10^{16},\,6.5\times10^{15},\,1.2\times10^{15}\}$ GeV, exceeding the required values by factors of $8.3\times10^5$, $8.7\times10^4$ and $1.2\times10^4$. Reaching the leptogenesis window therefore requires flavour vectors of order $10^{-3}$, three orders of magnitude below their generic size.}
\label{Figure 2}
\end{figure}

The hierarchy is therefore shared between the cycle volumes and the flavor vectors, and in the benchmark the latter dominates: the volumes supply a span $e^{-U_1}/e^{-U_3}\simeq21$, the flavor vectors supply $d_3^2/d_1^2\simeq71$, and the two partially cancel, leaving $M_3/M_1\simeq3.3$. Because the flavor vectors are here fixed by matching rather than computed from Eq. \eqref{dija}, the hierarchy is accommodated rather than predicted. Establishing the hierarchy as a prediction requires evaluating ${\rm Re}\,U_r$ from three-cycle volumes and $d_a^{(r)}$ from a specific compactification.

\section{Conclusions and discussions}
We have studied Majorana neutrino masses and the baryon asymmetry in Type IIA intersecting D6-brane models, where Euclidean D2-instantons provide a common origin for both: they generate the right-handed Majorana mass feeding the seesaw, and the dimension-5 Weinberg operator. Because the instanton action is a wrapped three-cycle volume, the suppression is geometric rather than tied to a small Standard Model coupling, and for $\mathrm{Sp}(2)$ instantons the flavor structure reduces to products of flavor vectors. We stress that in this paper these vectors and the instanton actions are not computed from a compactification; left-handed neutrino masses, mixing angles and right-handed neutrino masses are inputs, and the instanton relations determine what the flavor vectors would have to be.

Matching the measured mass-squared differences excludes the Weinberg-dominated case and selects the seesaw-dominated case with normal ordering. Relative to type-II seesaw, radiative, Dirac and inverse-seesaw constructions, the mechanism needs no extended Higgs sector, no extra symmetries and no artificially small couplings, and renders neutrinos unavoidably Majorana; relative to heterotic orbifolds and F-theory GUTs, the Type IIA setting realises the instanton-induced Majorana mass explicitly. Swampland criteria prune the available infrastructure rather than yielding a formula for $\eta_B$: the Distance Conjecture bounds the instanton action; the de Sitter and Trans-Planckian Censorship conjectures suppress the reheating temperature.

The benchmark of Section 6 reproduces the oscillation data, gives $\eta_B \simeq 8.64 \times10^{-11}$ via thermal leptogenesis and predicts $\langle m_{\beta\beta}\rangle=2.594$ meV. The observables depend only on the input spectrum and the Casas–Ibarra matrix $R$, so the example is a consistency check rather than a prediction. The scan of Section 7 shows that the mass hierarchy is likewise accommodated rather than predicted: only $0.89\%$ of the sampled space places the right-handed masses in the leptogenesis window, the required flavor vectors cluster near $10^{-3}$, and no $\mathcal{O}(1)$ configuration is viable.

Converting this into a prediction requires evaluating $\mathrm{Re}\,U_r$ from three-cycle volumes and the flavour vectors from the worldsheet correlators; a quantitative treatment of the moduli-driven channel is likewise left for future work. A single Type IIA vacuum can accommodate the origin of neutrino mass and the matter-antimatter asymmetry; whether it predicts them remains to be established by the geometry.

\appendix
\section{Details of numerical example}
In this appendix, we provide the details of the numerical example used in Section 6. The left-handed neutrino masses are taken as follows:
\begin{equation} \label{mL1}
m_1 = 9.998224 \times 10^{-4} \text{eV}, \sqrt{m_1}=0.03162 \text{eV}^{1/2},
\end{equation}
\begin{equation} \label{mL2}
m_2 = 8.6992929 \times 10^{-3} \text{eV}, \sqrt{m_2}=0.09327 \text{eV}^{1/2},
\end{equation}
\begin{equation} \label{mL3}
m_3 = 0.05049009 \text{eV}, \sqrt{m_3}=0.2247 \text{eV}^{1/2}.
\end{equation}
Therefore, we have
\begin{equation} \label{m2L}
m^2_1 = 9.9965 \times 10^{-7} \text{eV}^2, \, m^2_2 = 7.5678 \times 10^{-5} \text{eV}^2, \, m^2_3= 2.5492 \times 10^{-3} \text{eV}^2,
\end{equation}
\begin{equation} \label{deltam2L}
m^2_{21} = m^2_2 -m^2_1= 7.4678 \times 10^{-5} \text{eV}^2, \quad m^2_{31} = m^2_3 -m^2_1= 2.5482 \times 10^{-3} \text{eV}^2
\end{equation}
and
\begin{equation} \label{mt}
m_t = \sum^3_{i=1} m_i =0.0602 \text{eV}.
\end{equation}
The real parts of the moduli $\text{Re} U_i$ are chosen as:
\begin{equation} \label{ReU1}
e^{-\text{Re}U_1}=0.125, \quad \text{Re}U_1=2.07944,
\end{equation}
\begin{equation} \label{ReU2}
e^{-\text{Re}U_2}=0.0325, \quad \text{Re}U_2=3.4265,
\end{equation}
\begin{equation} \label{ReU3}
e^{-\text{Re}U_3}=0.0059, \quad \text{Re}U_3=5.1328.
\end{equation}
Based on experimental data, we select the following parameters:
\begin{equation} \label{thetadeltaalpha}
\theta_{12} =33.5^{\circ}, \, \theta_{23}=45^{\circ}, \, \theta_{13}= 8.5^{\circ}, \, \delta=234^{\circ}, \alpha_1 =0^{\circ}, \alpha_2=30^{\circ}.
\end{equation}
The resulting PMNS matrix is then given by $U= U_0 \cdot \text{diag} (1, e^{i15^{\circ}},1)$. 

Based on the definition of \eqref{UPMNS1}, we obtain the following values: $s_{12} =0.5519$, $c_{12}=0.8339$, $s_{23}=0.7071$, $c_{23}=0.7071$, $s_{13}=0.1478$, $c_{13}=0.9890$, $e^{-i\delta}=-0.5878+i0.8090$, $e^{i\delta}=-0.5878-i0.8090$, $e^{i15^{\circ}}=0.9659+i0.2588$. Consequently, the PMNS matrix takes the form
\begin{equation} \label{UPMNS4}
U_{PMNS} =
\left[ \begin{array}{ccc}		
0.8247, & 0.5272+i0.1413, & -0.0869+i0.1196 \\		
-0.3390+i0.0705, & 0.5902+i0.2065, & 0.6993 \\		
0.4414+i0.0705, & -0.5489-i0.0987, & 0.6993\\		
\end{array}	\right]
\end{equation}
We set $M_s=10^{17}\text{GeV}$ and consider the following right-handed neutrino masses:
\begin{equation} \label{MR1}
M_1 = 3.0 \times 10^{10} \text{GeV}= 3.0 \times 10^{19} \text{eV}, \sqrt{M_1}=5.477 \times 10^{9} \text{eV}^{1/2},
\end{equation}
\begin{equation} \label{MR2}
M_2 = 7.5 \times 10^{10} \text{GeV}= 7.5 \times 10^{19} \text{eV}, \sqrt{M_2}=8.660 \times 10^{9} \text{eV}^{1/2},
\end{equation}
\begin{equation} \label{MR3}
M_3 = 1.0 \times 10^{11} \text{GeV}= 1.0 \times 10^{20} \text{eV}, \sqrt{M_3}=1.0 \times 10^{10} \text{eV}^{1/2}
\end{equation}
For simplicity, we assume the matrix $d^{(r)}_a$ to be diagonal, which gives
\begin{equation} \label{d1}
d_1 = \sqrt{\frac{M_1}{2 M_s e^{-\text{Re}U_1}}} = 1.0954 \times 10^{-3},
\end{equation}
\begin{equation} \label{d2}
d_2 = \sqrt{\frac{M_2}{2 M_s e^{-\text{Re}U_2}}} = 3.397 \times 10^{-3},
\end{equation}
\begin{equation} \label{d3}
d_3 = \sqrt{\frac{M_3}{2 M_s e^{-\text{Re}U_3}}} = 9.206 \times 10^{-3}.
\end{equation}
In this work, we adopt the rotation matrix $R$ defined as
\begin{equation} \label{mR}
R
=\left[ \begin{array}{ccc}		
\cos z & 0 & \sin z \\		
0 & 1 & 0 \\		
-\sin z & 0 & \cos z\\		
\end{array}	\right]
\end{equation}
with $z= \frac{\pi}{4} + i0.25$. From \eqref{YnuCI}, we can define matrix $D$ 
\begin{equation} \label{mD}
D \equiv \sqrt{M^R} R \sqrt{m_{\nu}}
=\left[ \begin{array}{ccc}		
\sqrt{M_1} \cos z \sqrt{m_1} & 0 & \sqrt{M_1} \sin z \sqrt{m_3} \\		
0 & \sqrt{M_2} \sqrt{m_2} & 0 \\		
-\sqrt{M_3} \sin z \sqrt{m_1} & 0 & \sqrt{M_3} \cos z \sqrt{m_3}\\		
\end{array}	\right].
\end{equation}
As a result, we obtain the matrix elements of $D$:
\begin{equation} \label{D11}
D_{11} = \sqrt{M_1} \cos{z} \sqrt{m_1} = 1.263 \times 10^8 - i 3.093 \times 10^7, \, |D^2_{11}|=1.691 \times 10^{16};
\end{equation}
\begin{equation} \label{D13}
D_{13} = \sqrt{M_1} \sin{z} \sqrt{m_3} = 8.975 \times 10^8 + i 2.198 \times 10^8, \, |D^2_{13}|=8.53\times 10^{17};
\end{equation}
\begin{equation} \label{D22}
D_{22} = \sqrt{M_2} \sqrt{m_2} = 8.077 \times 10^8, \, |D^2_{22}|=6.524 \times 10^{17};
\end{equation}
\begin{equation} \label{D31}
D_{31} = -\sqrt{M_3} \sin{z} \sqrt{m_1} = -2.306 \times 10^8 - i 5.647 \times 10^7, \, |D^2_{31}|=5.6365 \times 10^{17};
\end{equation}
\begin{equation} \label{D33}
D_{33} = \sqrt{M_3} \cos{z} \sqrt{m_3} = 1.639 \times 10^9 - i 4.013 \times 10^8, \, |D^2_{33}|=2.847 \times 10^{18},
\end{equation}
and all other elements vanish.

We then compute two elements of $(Y_{\nu} Y^{\dagger}_{\nu})$:
\begin{equation} \label{YY13}
(Y_{\nu} Y^{\dagger}_{\nu})_{13} = \frac{1}{v^2} (D_{11} D^{\star}_{13} + D_{13} D^{\star}_{33}) = 4.479 \times 10^{-5} + i2.427 \times 10^{-5}
\end{equation}
\begin{equation} \label{YY132}
[(Y_{\nu} Y^{\dagger}_{\nu})_{13}]^2 = 1.417 \times 10^{-9} + i2.174 \times 10^{-9}, \, \text{Im} [(Y_{\nu} Y^{\dagger}_{\nu})^2_{13}] = 2.174 \times 10^{-9},
\end{equation}
and $(Y_{\nu} Y^{\dagger}_{\nu})_{12}=0$. 

Given our chosen values for $M_1$ and $M_3$, we obtain 
\begin{equation} \label{x}
x= \left( \frac{M^2_3}{M^2_1}\right) = \left( \frac{1.0 \times 10^{20}}{3.0 \times 10^{19}}\right)^2 = 11.11.
\end{equation}
Substituting this value of $x$ into \eqref{fMjM1} yields $f(11.11)=2.819$. Then, from \eqref{e1Ynu2}, we find 
\begin{equation} \label{epsilon1}
\epsilon_1= 1.546 \times 10^{-6}.
\end{equation}
The strength of washout is governed by the effective neutrino mass
\begin{equation} \label{tildem1}
\tilde{m}_1= \frac{(Y_{\nu} Y^{\dagger}_{\nu})_{11} v^2}{M_1} =0.029 \text{eV}.
\end{equation}
Using $m_{\star} = 1.1 \times 10^{-3}\text{eV}$, we further obtain
\begin{equation} \label{K}
K= \frac{\tilde{m}_1}{m_{\star}} =26.36,
\end{equation}
\begin{equation} \label{Kappa}
\kappa = \frac{0.3}{K (\ln K)^{0.6}}= 5.5886 \times 10^{-3},
\end{equation}
and
\begin{equation} \label{etaB}
\eta_B \approx 0.01 \kappa \epsilon_1 \approx 8.64 \times 10^{-11},
\end{equation}
within an order of magnitude of the observed value $\eta^{obs}_B =6.1  \times 10^{-10}$. 

We define $A= D U^{\dagger}$ and $Y_{\nu} = i A/v$, which leads to the following results:
\begin{equation} \label{A11}
A_{11} = (D U^{\dagger})_{11} = D_{11} U^{\dagger}_{11} + D_{13} U^{\dagger}_{31} = 5.250 \times 10^7 - i 1.519 \times 10^8,
\end{equation}
\begin{equation} \label{Ynu11}
Y^{\nu}_{11} = \frac{i}{v} A_{11} = 8.730 \times 10^{-4} + i 3.018 \times 10^{-4},
\end{equation}
\begin{equation} \label{A12}
A_{12} = (D U^{\dagger})_{12} = D_{11} U^{\dagger}_{12} + D_{13} U^{\dagger}_{32} = 5.826 \times 10^8 + i 1.553 \times 10^8,
\end{equation}
\begin{equation} \label{Ynu12}
Y^{\nu}_{12} = \frac{i}{v} A_{12} = -8.925 \times 10^{-4} + i 3.348 \times 10^{-3},
\end{equation}
\begin{equation} \label{A13}
A_{13} = (D U^{\dagger})_{13} = D_{11} U^{\dagger}_{13} + D_{13} U^{\dagger}_{33} = 6.812 \times 10^8 + i 1.311 \times 10^8,
\end{equation}
\begin{equation} \label{Ynu13}
Y^{\nu}_{13} = \frac{i}{v} A_{13} = -7.534 \times 10^{-4} + i 3.915 \times 10^{-3},
\end{equation}
\begin{equation} \label{A21}
A_{21} = (D U^{\dagger})_{21} = D_{22} U^{\dagger}_{21} = 4.258 \times 10^8 - i 1.141 \times 10^8,
\end{equation}
\begin{equation} \label{Ynu21}
Y^{\nu}_{21} = \frac{i}{v} A_{21} = 6.557 \times 10^{-4} + i 2.447 \times 10^{-3},
\end{equation}
\begin{equation} \label{A22}
A_{22} = (D U^{\dagger})_{22} = D_{22} U^{\dagger}_{22} = 4.767 \times 10^8 - i 1.668 \times 10^8,
\end{equation}
\begin{equation} \label{Ynu22}
Y^{\nu}_{22} = \frac{i}{v} A_{22} = 9.586 \times 10^{-4} + i 2.740 \times 10^{-3},
\end{equation}
\begin{equation} \label{A23}
A_{23} = (D U^{\dagger})_{23} = D_{22} U^{\dagger}_{23} = -4.433 \times 10^8 + i 7.972 \times 10^7,
\end{equation}
\begin{equation} \label{Ynu23}
Y^{\nu}_{23} = \frac{i}{v} A_{23} = -4.582 \times 10^{-4} - i 2.548 \times 10^{-3},
\end{equation}
\begin{equation} \label{A31}
A_{31} = (D U^{\dagger})_{31} = D_{31} U^{\dagger}_{11} + D_{33} U^{\dagger}_{31} = -3.806 \times 10^8 - i 2.078 \times 10^8,
\end{equation}
\begin{equation} \label{Ynu31}
Y^{\nu}_{31} = \frac{i}{v} A_{31} = 1.194 \times 10^{-3} - i 2.187 \times 10^{-3},
\end{equation}
\begin{equation} \label{A32}
A_{32} = (D U^{\dagger})_{32} = D_{31} U^{\dagger}_{12} + D_{33} U^{\dagger}_{32} = 1.220 \times 10^9 - i 2.452 \times 10^8,
\end{equation}
\begin{equation} \label{Ynu32}
Y^{\nu}_{32} = \frac{i}{v} A_{32} = 1.409 \times 10^{-3} + i 7.011 \times 10^{-3},
\end{equation}
\begin{equation} \label{A33}
A_{33} = (D U^{\dagger})_{33} = D_{31} U^{\dagger}_{13} + D_{33} U^{\dagger}_{33} = 1.040 \times 10^9 - i 2.893 \times 10^8,
\end{equation}
\begin{equation} \label{Ynu33}
Y^{\nu}_{33} = \frac{i}{v} A_{33} = 1.663 \times 10^{-3} + i 5.977 \times 10^{-3}.
\end{equation}
These results give the $Y^{\nu}$ matrix:
\begin{equation} \label{Ynu}
Y^{\nu} 
\approx 10^{-3} \times 
\left[ \begin{array}{ccc}		
0.873+i0.302, & -0.893+i3.348, & -0.753+i0.392 \\		
0.656+i2.447, & 0.959+i2.740, & -0.458-i2.548  \\		
1.194-i2.187, & 1.409+i7.011, & 1.663+i5.977\\		
\end{array}	\right]
\end{equation}
Besides, the effective Majorana mass is:
\begin{equation} \label{effneumass}
\langle m_{\beta\beta} \rangle \equiv |\sum^3_{i=1} U^2_{ei} m_i|= 2.594 \times 10^{-3} \text{eV} =2.594 \text{meV}.
\end{equation}

\acknowledgments
Thanks for the discussion with Zhong-Zhi Xianyu and Haipeng An. This work is supported by NSFC under Grants No. 12275146, the National Key R$\&$D Program of China (2021YFC2203100), the Dushi Program and the Shuimu Fellowship of Tsinghua University. For the purpose of open access, the authors have applied a CC BY public copyright licence to any Author Accepted Manuscript version arising.



\begin{thebibliography}{99}

\bibitem{fukuda1998evidence}
Y. Fukuda et al., \emph{Phys. Rev. Lett.} {81} (1998) 8.

\bibitem{ahmad2002direct}
Q. R. Ahmad et al., \emph{Phys. Rev. Lett.} {89} (2002) 1.

\bibitem{covi1996cp}
Laura Covi, Esteban Roulet, Francesco Vissani, \emph{Phys. Lett. B} {384} (1996) 169-174.

\bibitem{giudice2004towards}
G.F. Giudice, A. Notari, M. Raidal, A. Riotto, A. Strumia, \emph{Nucl. Phys. B} {685} (2004) 89-149.

\bibitem{Liu:2025zjs}
Yang Liu, \emph{Phys. Rev. D} {113} (2026) 066009.

\bibitem{ibanez2007neutrino}
L.E. Ibanez and A.M. Uranga, \emph{JHEP} {03} (2007) 052.

\bibitem{blumenhagen2007spacetime}
R. Blumenhagen, M. Cvetic and T. Weigand, \emph{Nucl. Phys. B} {771} (2007) 113.

\bibitem{cvetivc2007computation}
M. Cvetic, R. Richter and T. Weigand, \emph{Phys. Rev. D} {76} (2007) 086002.

\bibitem{ibanez2007instanton}
L.E. Ibanez, A.N. Schellekens and A.M. Uranga, \emph{JHEP} {06} (2007) 011.

\bibitem{Found:A}
Anil Ananthaswamy, \emph{SA Space and Physics} {2} (June 2019) 3.

\bibitem{antusch2007neutrino}
Stefan Antusch, Luis E. Ibanez and Tommaso Macri, \emph{JHEP} {09} (2007) 087.

\bibitem{junghans2019weakly}
Daniel Junghans, \emph{JHEP} {03} (2019) 150.

\bibitem{palti2019swampland}
Eran Palti, \emph{Fortschr. Phys.} {67} (2019) 1900037.

\bibitem{mandal2022toward}
Mandal, Miranda, Sanchez Garcia, Valle, Xu, \emph{Phys. Rev. D} {105} (2022) 095020.

\bibitem{cai2017trees}
Cai, Herrero-García, Schmidt, Vicente, Volkas, \emph{Front. Phys.} {5} (2017) 63.

\bibitem{behera2025phenomenology}
MK.Behera, P.Ittisamai, C.Pongkitivanichkul, P.Uttayarat, \emph{Eur. Phys. J. C} {85} (2025) 1316. 

\bibitem{vergados2016neutrinoless}
Vergados, Ejiri, Simkovic, \emph{IJMP E} {25} (2016) 1630007.

\bibitem{giedt2005massive}
Giedt, Kane, Langacker, Nelson, \emph{Phys. Rev. D} {71} (2005) 115013.

\bibitem{tatar2009right}
R.Tatar, Y.Tsuchiya, T Watari, \emph{Nucl. Phys. B} {823}.1-2 (2009) :1-46.

\bibitem{Bouchard:2009bu}
Bouchard, Heckman, Seo, Vafa, \emph{JHEP} {01} (2010) 061.

\bibitem{hamada2014flavor}
Y. Hamada, T. Kobayashi, S. Uemura, \emph{JHEP} {05} (2014) 116.

\bibitem{ooguri2007geometry}
H. Ooguri, C. Vafa, \emph{Nucl. Phys. B} {766} (2007) 21.

\bibitem{garg2019bounds}
SK Garg, C Krishnan, \emph{JHEP} {11} (2019) 075.

\bibitem{mohammadi2022brane}
A Mohammadi, T Golanbari, S Nasri, K Saaidi, \emph{Astropart. Phys.} {142} (2022) 102734.

\bibitem{king2024toward}
SF King, D Marfatia, MH Rahat, \emph{Phys.Rev.D} {109} (2024) 035014.

\bibitem{blumenhagen2005toward}
R. Blumenhagen, M. Cvetic, P. Langacker and G. Shiu, \emph{Annu. Rev. Nucl. Part. Sci.} {55} (2005) 71. 

\bibitem{cvetivc2010mass}
M. Cvetic, J. Halverson and R. Richter, \emph{JHEP} {07} (2010) 005. 

\bibitem{navas2024review}
Navas S, Amsler C, Gutsche T, et al., \emph{Phys. Rev. D} {110} (2024) 3.

\bibitem{planck2020planck}
Planck Collaboration (Aghanim et al.), \emph{Astron. Astrophys.} {641} (2020) A6.


\end{thebibliography}


\end{document}